\documentclass[acmsmall,screen]{acmart}
\AtBeginDocument{%
  \providecommand\BibTeX{{%
    \normalfont B\kern-0.5em{\scshape i\kern-0.25em b}\kern-0.8em\TeX}}}

\usepackage{natbib}
\usepackage{wrapfig}
\usepackage{booktabs}   %% For formal tables:
\usepackage{subcaption} %% For complex figures with subfigures/subcaptions
\usepackage{color}
\usepackage{colortbl}
\usepackage{cancel}
\usepackage{dashbox}
\usepackage[normalem]{ulem}
\usepackage{mathtools}
\usepackage{ textcomp }
\usepackage{makecell}
\usepackage{boldline}
\usepackage{tikz}
\newcommand{\stkout}[1]{\ifmmode\text{\sout{\ensuremath{#1}}}\else\sout{#1}\fi}
\definecolor{DOrange}{RGB}{145,20,20}
\definecolor{Red}{RGB}{255,0,0}
\definecolor{DeepGreen}{RGB}{5,102,8}
\definecolor{LGray}{gray}{0.9}
\definecolor{LCyan}{rgb}{0.88,1,1}

\usepackage{wrapfig}
\usepackage[frozencache,cachedir=minted-cache]{minted}
\usepackage{bbding}
\usepackage{listings} 
\usepackage{semantic}
\usepackage{tikz,pgf}
\usepackage{stmaryrd}
\usepackage{xcolor,colortbl}
\usepackage{ebproof}
\usepackage{relsize}
\setpremisesspace{1cm}

\usepackage{tikz}
\usepackage{xfrac} 
\usepackage{authorcomments}
\newcommand{\name}{\textsf{Hegel}}

\newcommand{\T}[1]{\mbox{$\xhookrightarrow{#1}$}}
\newcommand{\automata}{\textit{Qualified Tree Automata}\,}
\newcommand{\automaton}{\textit{Qualified Tree Automaton}\,}
\newcommand{\calculus}{\mbox{$\lambda_{\S{qta}}$\,}}

\usepackage{stmaryrd}

\definecolor{mygreen}{rgb}{0,0.6,0}
\definecolor{mygray}{rgb}{0.5,0.5,0.5}
\definecolor{mymauve}{rgb}{0.58,0,0.82}
\definecolor{amber}{rgb}{1.0, 0.75, 0.0}
\definecolor{bargreen}{rgb}{0.0, 0.5, 0.0}
\renewcommand{\S}[1]{\mbox{$\mathsf{#1}$}}

\lstnewenvironment{codetable}
  {\lstset{language=ML,
    basicstyle=\footnotesize,% basic font setting
    }%
  }
  {}
\usepackage{caption}

\tikzoption{base font}{\def\tikz@base@textfont{#1}}
\tikzset{
  base font=\sffamily,
}
\usepackage{amsmath}
\usepackage{supertabular}
\usepackage{geometry}
\usepackage{ifthen}
\usepackage{alltt}%hack
\usepackage{color}

\usepackage{proof}
\usepackage{multirow}
\usepackage{multicol}
\usepackage{semantic}
\usepackage{cancel}
\usepackage{algpseudocode}
\usepackage[noend,noline]{algorithm2e}
\usepackage{pgfplots, pgfplotstable}

\usepackage[normalem]{ulem}

\hypersetup{draft}
\begin{document}

\title{Close is Good Enough: Component-Based Synthesis Modulo Logical Similarity}

%Other : Decorating the Trees has its gifts : Qualified Tree Automata for searching in sparse spaces
%
% >>>>>>> 6eddf73ef4ccf7259a0cdd882a897007a3516ee8
%\author{Anonymous}
\author{Ashish Mishra}
\orcid{0000-0002-0513-3107}
\affiliation{
  \institution{IIT Hyderabad}            %% \institution is required
  \country{India}                    %% \country is recommended
}
\email{mishraashish@cse.iith.ac.in}

\author{Suresh Jagannathan}
\orcid{0000-0001-6871-2424}
\affiliation{
  \institution{Purdue University}            %% \institution is required
  \country{USA}                    %% \country is recommended
}
\email{suresh@cs.purdue.edu}
%% \authornote{Both authors contributed equally to this research.}
%% \email{trovato@corporation.com}
%% \orcid{1234-5678-9012}
%% \author{Anonymous}
%% \authornotemark[1]
%% \email{webmaster@marysville-ohio.com}
%% \affiliation{%
%%   \institution{Institute for Clarity in Documentation}
%%   \streetaddress{P.O. Box 1212}
%%   \city{Dublin}
%%   \state{Ohio}
%%   \country{USA}
%%   \postcode{43017-6221}
%% }

\renewcommand{\shortauthors}{Ashish Mishra, and Suresh Jagannathan}

\renewcommand{\ADD}[1]{#1}

\begin{abstract}
Component-based synthesis (CBS) aims to generate loop-free programs
from a set of libraries whose methods are annotated with specifications and whose output must satisfy a set of logical constraints, expressed as a query.  
The effectiveness of a CBS algorithm
critically depends on the severity of the constraints imposed by the
query. The more exact these constraints are, the sparser the space
of feasible solutions.  This maxim also applies when we enrich the
expressivity of the specifications affixed to library methods.  In both
cases, the search must now contend with constraints that may only hold
over a small number of the possible execution paths that can be
enumerated by a CBS procedure.

In this paper, we address this challenge by equipping CBS search with
the ability to reason about \emph{logical similarities} among the
paths it explores.  Our setting considers library methods equipped
with refinement-type specifications that enrich ordinary base types with a
set of rich logical qualifiers to constrain the set of values accepted
by that type.  We perform a search over a tree automata variant called
\emph{Qualified Tree Automata} that intelligently records information
about enumerated terms, leveraging subtyping constraints over the
refinement types associated with these terms to enable reasoning about
similarity among candidate solutions as search proceeds, thereby
avoiding exploration of semantically similar paths.

We present an implementation of this idea in a tool called \name\, and
provide a comprehensive evaluation that demonstrates \name's ability
to synthesize solutions to complex CBS queries that go well-beyond the
capabilities of the existing state-of-the-art.

\end{abstract}

%% \begin{CCSXML}
%% <ccs2012>
%%  <concept>
%%   <concept_id>10010520.10010553.10010562</concept_id>
%%   <concept_desc>Computer systems organization~Embedded systems</concept_desc>
%%   <concept_significance>500</concept_significance>
%%  </concept>
%%  <concept>
%%   <concept_id>10010520.10010575.10010755</concept_id>
%%   <concept_desc>Computer systems organization~Redundancy</concept_desc>
%%   <concept_significance>300</concept_significance>
%%  </concept>
%%  <concept>
%%   <concept_id>10010520.10010553.10010554</concept_id>
%%   <concept_desc>Computer systems organization~Robotics</concept_desc>
%%   <concept_significance>100</concept_significance>
%%  </concept>
%%  <concept>
%%   <concept_id>10003033.10003083.10003095</concept_id>
%%   <concept_desc>Networks~Network reliability</concept_desc>
%%   <concept_significance>100</concept_significance>
%%  </concept>
%% </ccs2012>
%% \end{CCSXML}

%% \ccsdesc[500]{Computer systems organization~Embedded systems}
%% \ccsdesc[300]{Computer systems organization~Redundancy}
%% \ccsdesc{Computer systems organization~Robotics}
%% \ccsdesc[100]{Networks~Network reliability}

%% %%
%% %% Keywords. The author(s) should pick words that accurately describe
%% %% the work being presented. Separate the keywords with commas.
%% \keywords{datasets, neural networks, gaze detection, text tagging}

\maketitle
\section{Introduction}
\label{sec:introduction}

Component-based synthesis (CBS) aims to generate loop-free programs
from a library of \emph{components}, typically defined as methods
provided by an API. At the heart of any CBS implementation is a
search problem over a hypothesis space of programs that ``glue''
components together using basic control primitives, such as
conditionals and function applications. If the attributes defining
the behavior of a component are not overly constrained, or when
queries are reasonably general, the search for a feasible solution can
be tractable. When this is not the case, however, the search can become
substantially more difficult because the number of feasible programs
that represent a solution is a much smaller fraction of the search
space.
  
Intuitively, we can define CBS search as a reachability analysis over
a graph that relates candidate methods based on their type or other
similar defining attributes.  For example, a node in this graph associated
with a method that has a particular result type can be connected to
any node corresponding to a method that accepts an argument of this
type. Such connections can be used by the synthesizer to produce a
candidate solution that sequences these methods together, yielding a
subgraph that connects input sources (e.g., function arguments) to
output targets (i.e., queries).

Prior work~\cite{sypet,tygus,ecta} has considered the construction of such graphs using simple type-based
specifications. In this paper, we propose to allow richer query
specifications in the form of refinement types~\cite{JV21} that both
decorate library methods and serve as the basis for synthesis
queries. Fortunately, advances in automated theorem proving have made
it increasingly common to have libraries be equipped with such rich
specifications~\cite{fstar,vocal}, and there is no reason to
believe that this trend will not continue to accelerate in the future,
making our setup both practical and topical. Indeed, the idea of using
refinement type specifications to guide a synthesis procedure has been
explored in a number of other recent
systems~\cite{synquid,cobalt-tech}; our focus on devising an efficient
CBS procedure that must contend with a sparse solution search space
differentiates our work in a number of significant ways from these
other efforts, as we describe below.

% \AM{Needs some reshaping for the new QTA, Highlighting that the new QTA is able to capture the space of refinement typed programs}

Devising an efficient CBS implementation in the presence of
fine-grained, complex specifications enabled by the use of refinement
types is challenging because the constraints defined by a type's
refinement may greatly restrict the set of feasible solutions. Simple
enumerative methods are, therefore, unlikely to be effective in this
setting. To address this challenge, we devise a novel tree automata
representation, called \automata (QTA), as an extension of finite tree
automata~\cite{tata} with constraints. A distinguishing feature of a
QTA is its support for logical implication constraints, which allows
us to identify semantically-related portions of the automata. In
particular, library methods and qualifiers in the method's refinement
type are treated as symbols for the QTA; implication constraints over
transitions allow modeling and reasoning about subtyping constraints
directly within the automata. Consequently, any program accepted by a
QTA is well-typed under the typing semantics of the refinement type
system.

%% We couple this FTA with an enumeration procedure that 
%%   constructs 
%%   a solution space, governed by the shape and semantics of
%% the query and the libraries available to the synthesizer. Central to
%% this construction is our exploitation of \textit{typing semantics} and \textit{logical similarities}
%% between transitions in the QTA to minimize duplicate or
%% unprofitable exploration.  

% \sout{Similar to prior work~\cite{ecta,Egg,VSA03,flashfill15,YRS23,blaze} that use tree
% automata-style data structures~\cite{tata} to represent the candidate
% search space, we propose to use tree automata to represent our
% enriched type graph of refinement-typed terms. To do this, we
% introduce \automata (QTA), 
% a tree automata representation tailored for
% efficiently managing the incremental construction of this space.
% Qualifiers in a library method's refinement type are attached to QTA
% states and reify the semantics of enumerated terms.}
% % Write about reductions

We leverage a QTA's structure to develop a CBS algorithm that tracks
both (i) irrelevant portions of the automata, i.e., portions of the
automata that do not correspond to well-typed terms, as well as (ii)
semantic similarities between terms, leveraging the use of a logical
subtyping relation during exploration to prune logically similar paths
during search. Our main insight is that the notion of intersection
available on finite tree automata naturally generalizes to a notion of
\textit{semantic} intersection over QTAs that can be exploited to
yield an efficient enumeration of the term space. We present two QTA
reduction procedures that concretize this intuition: a) a pruning
strategy to eliminate unproductive (sub)automata and b) a semantic
similarity mechanism that uses the type system's subtyping relation to
identify logically similar terms. These techniques enable efficient
exploration even when the set of feasible solutions is very small.

This paper makes the following contributions: 
\begin{itemize}
\item We present a new approach to address scalability and
  expressivity limitations in existing CBS frameworks, especially in
  the presence of expressive type-based queries that impose
  significant semantic constraints on the set of feasible solutions.
   
\item Our main insight entails directly embedding refinement-type
  specifications into a tree automata representation tailored to
  compactly represent sparse search spaces.
  
\item 
    We develop a specification-guided construction procedure of the
    search space used by the synthesizer using QTAs, instantiated with
    two reduction strategies to allow efficient pruning and compaction
    of the search space. We show that our algorithm is both sound and
    complete.

\item We present a detailed evaluation study using \name, a tool that
  incorporates these ideas and provides CBS capabilities for OCaml
  programs. Our results demonstrate the feasibility of efficiently
  synthesizing complex CBS queries even when the solution space is
  very sparse. These results support our claim that \name\ enables
  the solution of a variety of complex CBS problems that are outside
  the capabilities of existing approaches.

\end{itemize}

\noindent The remainder of the paper is organized as follows. In the
next section, we provide additional motivation and a detailed overview
of our synthesis approach. Section~\ref{sec:pftree} provides
background definitions. Section~\ref{sec:qta} formalizes QTAs.
Section~\ref{sec:synthesis} presents our synthesis
algorithm. Section~\ref{sec:reductions} discusses our QTA reduction
strategies. Soundness results are given in Section~\ref{sec:proofs}.
Details about the implementation, along with benchmark results, are
presented in Sections~\ref{sec:impl} and~\ref{sec:eval}. Related work
is given in Section~\ref{sec:related}, and conclusions are presented
in Section~\ref{sec:conc}.

\section{Motivation and Overview}
\label{motivation}

We motivate our approach using the synthesis problem shown in
Figure~\ref{fig:motivation}. The synthesizer takes two inputs. The
first is a library of OCaml functions specified using their
type signatures. Figure~\ref{fig:motivation}(a) shows a portion of
this library relevant to the example; it includes functions over
lists, integers, tuples, etc. For example, \S{splitAt} is a function
that takes an integer, a polymorphic list and returns a pair 
of polymorphic lists. For each function, we also present a
commented-out refinement type specification, which can be ignored for
the moment.

The second input to the synthesizer is a query, also represented as a
type signature. This is given the name \S{goal} in
Figure~\ref{fig:motivation}(b).  A solution to the query is a
synthesized OCaml function that, given two integers and a list, produces a
pair of lists of the same type as its argument list. The query in
this example is quite liberal in the solutions it admits.
Figure~\ref{fig:motivation}(c) shows two out of many possible
solutions.

% [basicstyle=\linespread{0.9}\small\sf,breaklines=true,language=ML]
{\small
\begin{figure*}
\begin{subfigure}[b]{0.50\textwidth}
\begin{minted}[fontsize = \footnotesize, mathescape=true, escapeinside=&&]{ocaml}
&\textcolor{blue}{(a) A part of the library}&

(*take : (x : nat) -> (xs : [a]) ->
{v : [a] $\mid$ len (v) $\leq$ x $\vee$ len (v) = 0}*)
val take : int -> [a] -> [a]

(*splitAt : (x : nat) -> (xs : [a]) ->
{v : (f : [a], s : [a]) $\mid$ len (f) $\leq$ x $\wedge$
(len (s) $\leq$ len (xs) - x)}*) 
val splitAt : int -> [a] -> ([a],[a])

(*incr : (x : nat) -> {v:nat $\mid$ x = x + 1}*)
val incr : int -> int

(*decr : (x : nat) -> {v:int $\mid$ x = x - 1}*)
val decr : int -> int

(*fst : (x : ([a], [a])) -> 
{v : [a] $\mid$ v = fst (x)}*)
val fst : ([a], [a]) -> [a]

(*snd : (x : ([a], [a])) -> 
{v : [a] $\mid$ v = snd (x)}*)
val snd : ([a], [a]) -> [a]

(*flatten : (ts : a tree) -> {v : [a] $\mid$ 
elems(v)=elems(ts)}*)
val flatten : a tree -> [a]
\end{minted}
%\caption{A list of library specifications.}
%\label{fig:lib}
\end{subfigure}
\begin{subfigure}[b]{0.45\textwidth}
\begin{minted}[fontsize = \footnotesize, mathescape=true, escapeinside=&&]{ocaml}
(*parse : {xs : [a] $\mid$ even_len(xs)} -> 
{v : a tree $\mid$ elems(v)=elems(xs)}*)
val parse : [a] -> a tree 

(*clear : (xs : [a]) -> {v: [a] $\mid$ len(v) = 0}*)
val clear : [a] -> [a]

(*drop : (x : nat) -> (xs : [a]) ->
{v : [a] $\mid$ len (v) $\leq$ len (xs) - x}}*)
val drop : int -> [a] -> [a]

(*rev : (xs : [a]) -> {v : [a] $\mid$ len(v)=len(xs) $\wedge$ 
$\forall$ u, w. ord (u, w, xs) => ord (w, u, v)}*)
val rev : [a] -> [a]


&\textcolor{blue}{(b) A functional query type}&
(*goal : (x:nat) 
-> (y : nat) -> (z : [a]) 
-> { v : (f : [a], s : [a]) $\mid$ len (f) $\leq$ x 
$\wedge$ (len (s) $\leq$ len (z) - y}*)
goal : int -> int -> [a] -> ([a], [a])

&\textcolor{blue}{(c) A few correct solutions}&
 (*fun x y z -> 
 ((take x 
    (fst (splitAt y z))), 
  snd (splitAt y z) )*)
 &fun& x y z -> splitAt y (drop x z)
 &fun& x y z -> splitAt x (take y z)
 &$\ldots$&

\end{minted}
%\caption{Functional query-spec and a solution.}
%\label{fig:refined-synthesis}
\end{subfigure}
\caption{Motivating Example.}
\label{fig:motivation}
\end{figure*}
}
% present a few solutions

% There are many possible interesting solutions for the given query, some of which are listed below.
% \begin{minted}[mathescape=true,escapeinside=&&]{ocaml}
% &$\lambda$& x y z -> splitAt x (drop y z)
% &$\lambda$& x y z -> splitAt y (drop x z)
% &$\ldots$&
% \end{minted}

%Discuss how sota will find a solutions
To compute these solutions, the task of a component-based synthesizer
is to \textit{search} for a valid composition of functions found in
the library, satisfying the type signature of each function such that
the type of the composition aligns with the query type. A standard
approach to solving CBS problems involves treating search as a kind of
graph reachability problem in which a graph constructed over the types
of library functions (henceforth type graph) is explored to find
reachable paths between arguments and return
types~\cite{sypet,tygus,ecta}.

\begin{figure*}
\begin{subfigure}[b]{0.50\textwidth}
\centering
\advance\leftskip-1cm
\centering 
\includegraphics[scale=.40]{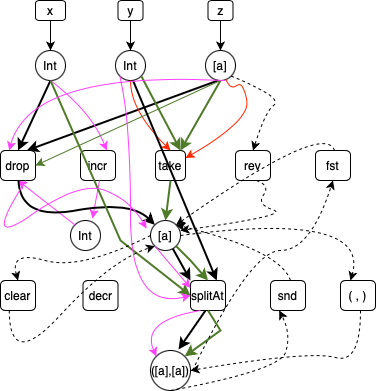}
 \caption{Types graph for the original example. }
 \label{fig:enumeration-o}
 \end{subfigure}
\begin{subfigure}[b]{0.40\textwidth}
\centering
\advance\leftskip-1cm
\centering 
\includegraphics[scale=.40]{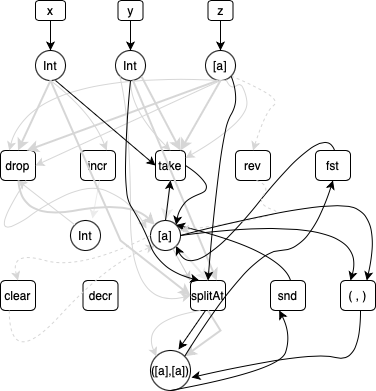}
 \caption{Types graph for the refined example.}
 \label{fig:enumeration-r}
 \end{subfigure}
\caption{Paths in the library for the original and refined query. We
  show multiple nodes for a type for ease of understanding.}
\end{figure*}

Figure~\ref{fig:enumeration-o} depicts this idea. We show the query
arguments' types as source nodes and the required goal type ([a],
[a]) as the target node. Edges connect ``producers'' and
``consumers'' of a type. \ADD{We have replicated the nodes for some of the types like \S{int}, and \S{[a]} for the clarity of presentation. The figure shows multiple possible paths
within the graph using library functions as intermediate transition
nodes.}  Paths are color-coded, each representing a distinct candidate
solution that leads from an input argument (e.g., \S{x,y} or \S{z}) to
the target. For instance, the black solid edges represent the first
solution shown in Figure~\ref{fig:motivation}(c). Similarly, the green
paths represent the solution \S{splitAt \ x \ (take\ y\ z)}.
The pink path represents the solution \S{splitAt \ y \ (drop\ (incr \ x))\ z}. 

Dashed edges show other feasible paths in the graph, not all of which
lead to a complete solution. Note that many paths have cycles and can
thus represent a set of potential solutions. In this example, since
there are multiple paths that lead to the target, a synthesis tool can
easily find one and return it as a solution. Indeed, when tested
with two state-of-the-art type-guided CBS tools~\cite{tygus} and~\cite{ecta}, a
correct solution was generated in less than 5 seconds.
% introduce a sparse search space
% >> However, as the queries become more complicated, simple types need to be augmented with additional structure/semantics.
\paragraph{Type-based CBS on refined queries}
Now, consider a slight variation of the query that refines the desired
solution by also requiring that in the returned pair, (i) the size of
the first list is less than or equal to the first argument and (ii)
the size of the second list is less than or equal to the original list
length minus the second argument.  \ADD{One way to express this query
  is through the use of method predicates or type qualifiers like \S{len}, \S{fst, snd},
  etc. to capture properties such as \textit{length} of the list,
  \textit{first} and \textit{second} projections of a tuple, etc.  The
  query represented in this way is given as the commented signature
  in Figure~\ref{fig:motivation}(b).}

Traditional enumerative exploration for this significantly more
constrained query is unlikely to be successful for two fairly apparent
reasons. First, the solution space, which was quite dense earlier,
reflecting the fact that most paths chosen by the synthesizer for
exploration would reach the target is now very sparse.
Figure~\ref{fig:enumeration-r} shows the new type graph for this
refined query. Notice that most of the earlier solutions shown in
different colors are now invalid and greyed-out. Only the paths in
solid back represent a feasible solution. The corresponding correct
term is given in the comment in Figure~\ref{fig:motivation}(c).
Second, the (base) type specifications associated with the libraries
are too weak to meaningfully guide the synthesizer toward these sparse
solutions or to filter out incorrect solutions. In particular, simple
types are incapable of differentiating the correct (black) path from
other previously-seen invalidated paths.

%\AM{The above section sets up the need for a efficient search in Refinement type space very clearly }
\paragraph{Exploration over an augmented search space}
What we require is the ability to add additional structure to the
graph beyond just the simple type signature that we currently have, to
allow infeasible paths to be detected and pruned, and
semantically-equivalent paths to be identified. Unfortunately,
incorporating such additions comes at a cost both in terms of graph
size and search complexity. To illustrate the issue with the former,
note that a single type node \textsf{int} in the original graph may
now expand into many nodes when qualified with semantic information
that capture more refined properties, e.g., positive \S{int}, negative
\S{int}, \S{int\ less\ than\ v}, etc. In fact, this set is unbounded
in general. The issue with the latter directly impacts how the
synthesizer is engineered. Existing
techniques~\cite{tygus,ecta,sypet} search over a pre-built, fully
expanded type graph of some size for the complete library coupled with
a pruning mechanism over this graph. However, building such graphs in
the presence of refined semantic information raises obvious
scalability concerns. To address these issues requires a new graph
representation that can concisely represent the space terms with
refined specifications and a new search procedure tailored to operate
over this new representation.

\subsection{Approach}

% \SJS{Move the Syntehsis problem definition here and make it explicitly expose the cbs over refinement types, by explicitly showing refinement types as part of the statement.}

% Synthesis 

% \TODO{Incorporate all details asked by Reviewer C.}

\subsubsection{\NEW{Synthesis Problem over Refinement Typed Libraries}}
\NEW{A component-based synthesis problem over a refinement-type annotated library can be thus defined formally as follows:}
 \NEW{
\paragraph{\bf Synthesis Problem}
Given a type environment $\Gamma$ that relates library functions $f_i = \lambda (\overline{x_{i,j}}). e_{f_i}$ with their refinement types
$f_i : \overline{(x_{i,j} : \tau_{i,j})} \rightarrow \{ \nu :  t_i \mid \phi_{i} \} \in \Gamma$, and a synthesis query  $\varphi = \overline{(y_i : \tau_i)} \rightarrow \{\nu : t \mid \phi \}$, a solution to a CBS problem seeks to synthesize an expression $e$, possibly using $f_i$, such that $\Gamma \vdash e : \varphi$ holds.  
}
% \NEW{Where $\Gamma \vdash e : \varphi$~\footnote{see supplemental material, \TODO{Need this definition.}} captures typing semantics for \calculus{}.}

\subsubsection{Compact Representation of the Search Space}
\NEW{A primary requirement towards solving the above problem is to compactly represent the space of 
well-typed terms.
There are several data structure options that we might choose from  that can serve this purpose.  Version Space Algebras~\cite{vsa},
e-graphs~\cite{egg-synthesis} and Finite Tree Automata
(FTA)~\cite{tata} are three well-studied examples. In particular, FTAs have been shown to be
effective in representing the space of untyped programs and allowing efficient search over them, satisfying a set of input-output
examples~\cite{Dillig23}. Furthermore, extensions of FTA with
equality constraints, dubbed
ECTAs~\cite{ecta, tata}, have been shown to be a useful
representation to represent simply-typed programs.  }

\NEW{Unfortunately, these approaches are ineffective in representing the
space of programs with refined specifications, such as those
considered in our motivating example.
For instance,  VSA and standard FTAs lack the ability to relate subprograms, while constrained FTAs~\cite{tata,ecta}, which allow \emph{syntactic} equality constraints between subterms, are insufficient when \emph{logical} equality/implication constraints are required. This requirement is clearly seen in the refined variant of our motivating example where the synthesis query establishes non-trivial semantic relationships between the list elements in the output.~\footnote{see supplemental material for a detailed comparison and limitation of these structures.}} 

\paragraph{Solution: Qualified Tree Automata}

\begin{wrapfigure}{r}{.25\textwidth}
\includegraphics[scale=.450]{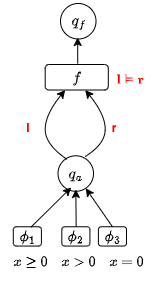}
\caption{A simple QTA for \{ $f (\phi_1 , \phi_2) |  \phi_1, \phi_2 \in \S{FOL}, \phi_1, \implies \phi_2 $ \}} 
\label{fig:compare-single}
\end{wrapfigure}
\NEW{To address these limitations, we introduce a new data structure, a \automata{} (QTA), that allows us to capture such logical constraints.
A QTA supports a richer alphabet than other FTA variants by incorporating logical qualifiers from decidable first-order theory fragments. While it also allows
constraints on its transitions similar to other constrained tree
automata~\cite{tata,ecta,cata}, it additionally supports semantic reasoning over these constraints (e.g., logical entailment),
rather than being limited to syntactic reasoning using equality or
dis-equality constraints.}

% \AM{I want to keep this as it just shows a QTA without any type information, thus answering Reviewers point about explain how it is different than other approaches. Added a line of explanation.}
\NEW{For instance, Figure~\ref{fig:compare-single} presents a QTA that captures the space of terms represented by the following sentence \{ $f (t_1 , t_2)\ |\ t_1, t_2 \in \{ \phi_1, \phi_2, \phi_3 \}$ $\wedge \ t_1 \implies t_2$ \} 
where both sub-trees are constrained using a \textit{logical entailment} constraint (defined later) on the transition 
%\SJ{This notation hasn't been introduced - is it necessary? If so, it needs to be explained.  }
(\textcolor{red}{{\sf l} $\vDash ${\sf r}}). Here, \textcolor{red}{l} and \textcolor{red}{r} are variables that capture a specific location in the automata and the constraint restricts which choice of $t_1$ and $t_2$ are acceptable.}
% Consequently, the QTA will only accept terms of the form $\{f (f_i, f_j)\ \mid\ f_i = f_j$\}. The symbols in the QTA also allows logical formulas $\phi_i$, from a decidable logic family.}

\NEW{These formulae and the constraints on the transition together restricts the language accepted by the QTA. In this example, the automata accepts terms $f (\phi_2, \phi_1)$ and $f (\phi_3, \phi_1)$ since, in both cases, the constraint $\phi_i \implies \phi_j$ holds but the QTA rejects other syntactically valid terms like $f (\phi_1, \phi_2)$ and $f (\phi_1, \phi_3)$.}
% \SJ{I would remove the following along with the figure.}
% The rightmost automaton in Figure~\ref{fig:comparison} shows a QTA for the modified example with logical constraints, the language accepted by the QTA includes $f (\phi_2, \phi_1)$ and $f (\phi_3, \phi_1)$ as in both cases the constraint $\phi_i \implies \phi_j$ holds but will reject other syntactically valid terms like $f (\phi_1, \phi_2)$ and $f (\phi_1, \phi_3)$.

% \hline
% \AM{How to relate this to the next paragraph?}

% This is because these
% richer specifications not only use constant base types like
% \textsf{int}, \textsf{char}, \textsf{bool}, etc., but also allow types
% to be qualified with logical formula (aka refinements).  
% \AM{The following paragraph needs some rewriting for clarity.}

%\AM{Added the below explicit paragraph heading, the text remains similar to the earlier version.}
% \hline
\paragraph{ Embedding Refinement-typed space using QTA}

Unlike traditional typed-program space embedding where  syntactic constraints are sufficient to compare base types, the space of well-typed refinement specifications not only use constant base types like
\textsf{int}, \textsf{char}, \textsf{bool}, etc., but also allow types
to be qualified with logical formula (aka refinements)~\cite{liquidoriginal}.
QTA are effective in expressing these structures and enable a compact embedding of these type structures.

\begin{wrapfigure}{r}{.20\textwidth}
\includegraphics[scale=.50]{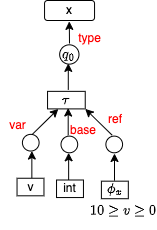}
\caption{A simple QTA for binding (x : \{ $\nu$ : int $\mid$ $10 \geq \nu \geq 0$ \})} 
\label{fig:baseqta}
\end{wrapfigure}

% \sout{A QTA supports a richer alphabet than prior work using FTAs, capturing
% program variables, base types, logical qualifiers, as well as
% \textit{refinement types}. Additionally, it also allows
% constraints on its transitions, similar to other constrained tree
% automata~\cite{tata,ecta,cata}, it provides the ability to perform
% semantic reasoning over these constraints (e.g., logical entailment),
% rather than being limited to reasoning over just equality or
% dis-equality constraints, as is the case with existing approaches.}

\NEW{To illustrate such an embedding, Figure~\ref{fig:baseqta} shows a transition in a
trivial QTA, representing a variable \S{x} having a type \{ $\nu$ :
int $\mid$ $10 \geq \nu \geq 0$ \}.  
States are shown as circles, and
transitions as rectangles. A transition can have zero or more
incoming states, with each having a \textit{position} label, shown over incoming arrows in transitions in \textcolor{red}{red}. 
For instance, the transition (\textcircled{$q_0$} $\xrightarrow{\textcolor{red}{\S{type}}}$ \fbox{\S{x}}) is used to capture  a standard type binding (x : $\tau$). 
These types can be \textit{refined},  by allowing
transitions for an alphabet symbol $\tau$ (drawn from the set of base refinement
types). $\tau$ is a ternary symbol, with three incoming states (incoming arrows from states)
corresponding to the parameters, \textit{variable}, \textit{base-type} and a
\textit{logical formula} (i.e., refinement). The states themselves have incoming
transitions, and so on. 
In this example, the automata rooted at $q_0$ corresponds to the
  refinement type \{ $\nu$ : int $\mid$ $10 \geq \nu \geq 0$ \}.}

% \AM{This confused two of the reviewers, now we do not need this.}
% \sout{To precisely
% capture the space of well refinement-typed programs, an automata representation
% must, therefore, also admit these structures. More significantly, to
% effectively reason about qualified terms within an automata-based
% framework additionally requires transplanting notions of logical
% entailment used in the type system as constraints on transitions in
% the automata. For instance, Figure~\ref{fig:explosion} shows some
% possible terms, involving functions \S{take}, \S{incr}, and \S{clear},
% along with their refinement types, that a synthesizer may have to
% enumerate.}

% \sout{We address this issue in two ways: First, to compactly represent the
% augmented space, we propose a novel variant of Finite Tree
% Automata~\cite{tata}, called \automata{} (QTA), discussed in detail
% below. Second, we define our synthesis procedure to exploit
% opportunities to \textit{eagerly} prune infeasible as well as
% redundant portions of the search space while still maintaining
% completeness of the enumeration procedure.
% }

% \SJS{Here is where we define the variant of the problem as by now they have seen the definition of QTA}
\subsubsection{Efficient Enumeration}
\begin{figure}
\includegraphics[scale=.450]{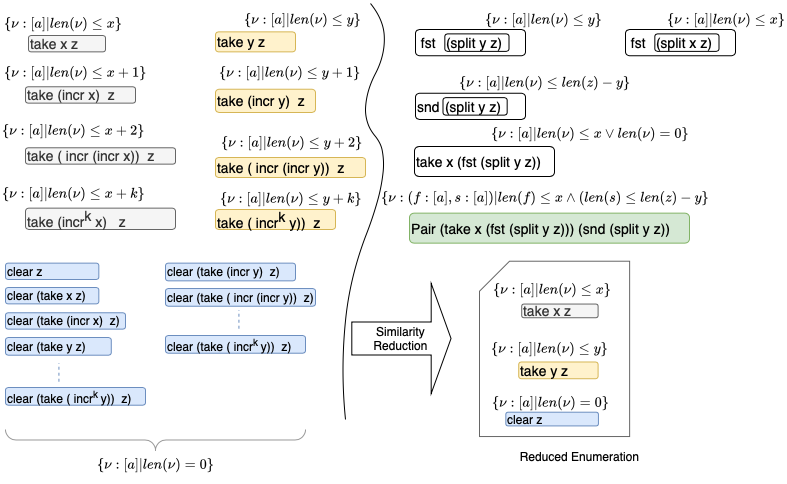}
 \caption{A partial set of terms in the augmented type graph for the
   example along with their augmented refinement type. The curved line
   partitions the terms which lead to a solution (shown in green)
   towards the right (top) and other explored terms which although
   type-correct do not lead to a solution (left). The right bottom
   shows the list of terms produced by \name{}.}
 \label{fig:explosion}
\end{figure}

Compactly representing the program space is only part of the solution, however.  Another
significant challenge for CBS is devising an efficient enumeration of
the program space. \NEW{For instance, Figure~\ref{fig:explosion} shows some
possible terms, involving functions \S{take}, \S{incr}, and \S{clear},
along with their refinement types, that a synthesizer may have to
enumerate.} In our setting, na\"{\i}ve enumeration is not
feasible due to the sparseness of the solution space. In
Figure~\ref{fig:explosion}, for instance, only a fraction of the
overall space, specifically the terms on the right top of the curve in
the figure, are actually relevant to the solution. The terms on the
left of the curve in Figure~\ref{fig:explosion}, although type
correct, are irrelevant to the result.

Furthermore, refinement type information alone is often insufficient
to efficiently guide the synthesizer to a possible solution. For
example, in the figure, there is little guidance available to
determine that the terms on the left of the curve are unlikely to
contribute to a solution compared to the terms given on the right.
The cost of enumerating these ineffective terms becomes problematic as
larger terms are built from them.
\NEW{To address the enumeration challenge, our synthesis procedure exploits
opportunities to \textit{eagerly} prune infeasible as well as
redundant portions of the search space while still ensuring that the enumeration procedure is complete.}

\emph{Eager equivalence reduction}: This enumeration cost can be mitigated if we
can somehow prune out infeasible and redundant terms during the
enumeration process. Notice in Figure~\ref{fig:explosion} that
many of these terms, although syntactically different, have the same
(or equivalent) refinement type. For example, consider all the
\S{clear} function calls, which are shown in blue in the lower left of
the figure. These are all distinct terms, in total ($2k + 1$) in
number for any $k$ \S{incr} calls, but each has the same type, (\{
$\nu : [\S{a}] \mid len (\nu) = 0$ \}). We utilize these
type-equivalences to prune out all but one of these terms. The
high-level intuition is that for any given query and library, all or
none of these nodes will lead to a solution, so exploring along any
one is sufficient to reach a solution if one exists using them. For
instance, all the blue terms can be replaced with a single term
(\S{clear \ z}) shown in the lower right portion of the figure, showing
the reduced space. We might think that such equivalences occur only
between terms with overlapping structures, like all the terms in blue.
However, note that this equivalence reduction can be generalized to
any arbitrary pair of terms, e.g., see (\S{take \ x \ z}) on the left
of the curve, and (\S{fst\ (splitAt\ x\ z)}) on the right.

\emph{Similarity, rather than equivalence}: Tracking precise
equivalences, although useful as shown above, allows only a marginal
reduction of the augmented space. To see why, consider all the terms
in Figure~\ref{fig:explosion} in the upper left corner shown in the
grey and yellow colored boxes.

\begin{wrapfigure}{r}{.40\textwidth}
\includegraphics[scale=.400]{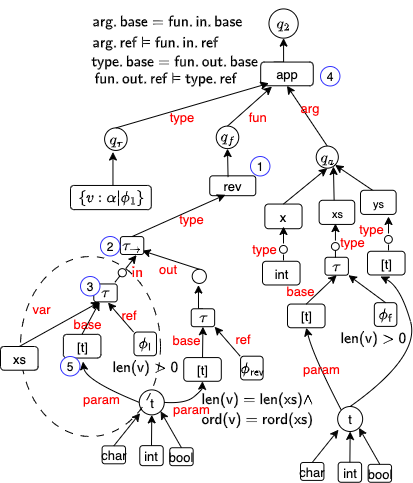}
\caption{A partially-defined QTA for our motivating example.  States are shown as circles, and transitions as rectangles. Some unimportant states and transitions are omitted for clarity. A transition can have zero or more incoming states, with each having a label denoting its
  \textit{position}, shown in \textcolor{red}{red}. Transitions can also have
  constraints over these positions. \textcircled{i} are labels used for elucidation.} 
\label{fig:introduceqta}
\end{wrapfigure}
% \begin{wrapfigure}{r}{.50\textwidth}
%     \includegraphics[scale=.50]{fig/ref-motivate-reduced.png}
%     \caption{A reduced QTA for the QTA in Figure~\ref{fig:refmotivate}}
%      \label{fig:refmotivate-reduced}
%  \end{wrapfigure}
Because these terms have valid refinement types, they are candidates
for enumeration by the synthesis search procedure. However, none of
these terms actually contribute to the solution. Unfortunately, exact
equivalence tracking, as described earlier, is ineffective in pruning
these terms since their type annotations are all distinct. Observe,
however, that the types of these terms are related to one another
under a refinement \textit{subtyping} relation. For example, the
refinement type for the first term in grey (\S{take \ x \ z}), is a
subtype of the refinement types of all other grey terms. Similarly,
the type of the first yellow term is a subtype of the type of all
other yellow terms. By defining a weaker \textit{similarity} relation
between terms based on the subtype relation, we can avoid exploring
possibly non-equivalent but \textit{similar} terms without affecting
the completeness of the search procedure. The use of a subtyping
relation as a proxy for similarity allows us to collapse, for example,
all grey and yellow terms into two representative candidates, as
depicted in the bottom-right of the figure.

Our main insight is to map the similarities and equivalences between
transitions in automata to the subtyping relation in the type
system. This allows us to use operations available on QTAs like
intersection and union to prune all but the most specific term (i.e.,
term with the lowest type in the subtype ordering).\\

\emph{Irrelevant candidate elimination}: Another thing to notice in
the example is that many functions in our example library, e.g.,
\S{parse}, \S{flatten}, etc. are irrelevant, as there are no available
arguments of the required type in the query or other library methods
to invoke them. This can happen due to typing restrictions on the
arguments, for example, the \S{parse} function requires that length of
the input list to be `even'. Or, it can happen simply because there are
no terms of the required type in the library or the terms that have
been enumerated. Pruning away such functions early on allows us to
avoid not only reducing the candidate space but potentially avoids
the need to enumerate many other larger terms built using them for
example, \S{(parse \ (flatten \ldots))}, \S{(clear \ (flatten \ (parse
  \ldots)))}, \S{(take \ x \ (flatten \ (parse \ldots)))}, \S{(take
  \ y \ (flatten \ (parse \ldots)))}, etc.

However, \NEW{a na\"{\i}ve enumeration would generate potentially many irrelevant functions and terms, not all of which are even well-typed. Efficiently filtering these terms is likely to be infeasible, in general, given the enumerative structure of the approach, and the cost of performing type checks, both of which involve an SMT query.}
% \sout{is
% non-trivial, as in many cases, this requires an analysis comparing the
% refinement types of each term in the library and enumerated terms,
% before discarding them.}  
Our next insight is that we can use
operations over the QTA to prune out portions of the QTA representing
such terms.

\subsubsection{QTA Through Examples}

%  \begin{figure}[htbp]
% \small
% % \begin{subfigure}[b]{0.40\textwidth}
% % \includegraphics[scale=.450]{fig/run1.png}
% % %\ottmetavars\\[0pt]

% % \subcaption{A small QTA for valid subtyping pairs between base refinement types.}
% % \label{fig:qtaruna}
% % \end{subfigure}
% \begin{subfigure}[b]{0.25\textwidth}
% \includegraphics[scale=.50]{fig/base_qta.png}
% \caption{A simple QTA for binding (x : \{ $\nu$ : int $\mid$ $10 \geq \nu \geq 0$ \})} 
% \label{fig:baseqta}
% \end{subfigure}
% \begin{subfigure}[b]{.55\textwidth}
% \includegraphics[scale=.450]{fig/introduceqta.png}
% \subcaption{A partial QTA for the Motivating example.} 
% \label{fig:introduceqta}
% \end{subfigure}
% \caption{QTA through examples: States are shown as circles, and transitions as rectangles. Some
%   unimportant states and transitions ommitted for clarity. A transition can have zero
%   or more incoming states, with each having a label denoting its
%   \textit{position}, shown in \textcolor{red}{red}. Transitions can also have
%   constraints over these positions. \textcircled{i} are labels used for elucidation.}
% \label{fig:qtarun} 
% \end{figure}

We highlight some of the details of \automata{} using a detailed example.

\NEW{Consider
Figure~\ref{fig:introduceqta} that depicts a portion of the QTA for our
motivating example, showing the annotated library function \S{rev} as a transition \fbox{\S{rev}} (we have used labels like \textcircled{1} for elucidation), along with its refinement type as another transition (\fbox{$\tau_{\rightarrow}$}) (\textcircled{2}). Note that because this is a function type, it has two children, one for its input \textit{argument} type and another for its \textit{return} type, with location labels \textcolor{red}{in} and \textcolor{red}{out} respectively. 
The argument's type is represented in the usual way,  as shown in Figure~\ref{fig:baseqta} (the dashed circle at \textcircled{3}). 
Transition $\fbox{\S{rev}}$ also has an outgoing edge to state $q_f$, representing the set of unary functions.
In general, for each $n-$ary function symbol, under $q_f$,
the \textcolor{red}{in} and \textcolor{red}{out} positions shows the sub-automata
modeling the argument(s) and result type for the function.}

\NEW{QTAs also support polymorphic refinement types. For instance, a polymorphic list type \S{[a]} is represented by a transition with list constructor \S{[t]} (\textcircled{5}) with a type parameter \S{t} and another automata for all  base type \S{t} that can be present on the incoming edge representing the constructor's type parameter.
}

\NEW{A QTA also efficiently captures refinement typing semantics.  For instance, transition (\fbox{\S{app}})(\textcircled{4}) captures the expected type application rule. It has three incoming states, $q_f$, representing a set of unary functions, $q_a$, a set of possible variables (along with their types) as arguments ((\S{x : int}), (\S{xs} :
  \{ [a] | len ($\nu$) > 0\}) and (\S{ys} : [a])), and $q_{\tau}$ that represent the inferred type for the application. 
Application typing constraints are added as constraints to the \S{app} transition, which captures not only equality for base types, 
function return types, the return type for the application
term (\S{type.base = fun.out.base}), the function input type, and
the argument base type, but also allows constraints with logical
entailment $\vDash$ and other constraints generated through logical
connectives like $\wedge$, $\vee$ etc. Thus, $\mathsf{arg.ref
  \vDash fun.in.ref}$ establishes a logical entailment constraint between
a function's formal and actual refinements.
}

\section{Preliminaries}
\label{sec:pftree}
% \AM{Make this very abstract, highlighting just two features of the QTA, exposing the specs and similarity reduction using subtyping information.}
\subsection{Background: Finite Tree Automata (FTA)}

A ranked alphabet ($\mathcal{F}$, \textsf{Arity} : $\mathcal{F}
\mapsto \mathbb{N}$) is a pair consisting of a set of symbols
($\mathcal{F}$), and a map that relates these symbols to their arity.
In the context of CBS, these symbols are those found in expressions,
library functions provided by the synthesis problem, query arguments, and associated types and refinements for each expression. For
example, constants and variables are symbols of arity zero, as are
constructors without arguments like \S{Nil}, etc. These form a set
$\mathcal{F}_0$.  A \textit{base refinement type} is represented as a
symbol {$\tau$} with arity 3 as shown in in
Figure~\ref{fig:motivation}. In general, the set of all symbols of
arity \textit{n} is denoted as $\mathcal{F}_n$.

\begin{definition}[Finite Tree Automata]
A bottom-up finite tree automaton (FTA),
$\mathcal{A}$ over a signature of ranked alphabet ($\mathcal{F}$, Arity : $\mathcal{F} \mapsto \mathbb{N}$ is a tuple ($Q$, $Q_f$, $\mathcal{F}$,
$\Delta$) where $Q$ is a set of states, $Q_f \subseteq Q$ is a set of
final states and $\Delta$ is a set of transition rules of the
following form : $f(q_1(x1), . . . , q_n(xn)) \rightarrow q(f(x1,
. . . , xn))$, where {\sf n} $\geq$ 0, $f \in \mathcal{F}_n$, $q, q_1, . . . ,
q_n \in Q$, $x_1, . . . , x_n \in \mathcal{X}$. where $\mathcal{X}$ is a
set of variables (symbols with arity 0).
\end{definition}    

\begin{example}
\label{ex:example-fta}
Let the signature over which a tree automata is defined be given by a
pair of symbols $\mathcal{F}$ = \{$f$, $g$, $a$\} and Arity = \{ $f
\mapsto$ 2, $g \mapsto$ 1, $a \mapsto$ 0\}. The set of states is $Q$
= \{ $q_a, q_g, q_f$ \}, let the final state be $q_f \in Q_f$ and let
$\Delta$ be given by the following transition rules:
\[
 \{a \rightarrow q_a(a) \ ; \  g(q_a(x)) \rightarrow q_g (g(x))  \ ; \ g(q_g(x)) \rightarrow q_g(g(x)) \ ; \ f(q_g(x), q_g(y)) \rightarrow q_f(f(x, y))\}
\]
\end{example}

Analogous to finite automata that accept strings over an alphabet of
characters, an FTA accepts trees, which are terms over $\mathcal{F}$. In
the example given above, the FTA accepts all valid trees of the form
$f (...)$, generated using functions $f, g$ and the constant $a$.

\subsection{Synthesis Language, \calculus}

% \AM{I have added the details of A-Normal form of the \calculus later in a separate section about details of synthesis, if needed it can be moved here.}

The target language of our synthesizer is a standard
A-normalized~\cite{flanagan} call-by-value typed  $\lambda$-calculus with
constructors, constants and variables, conditional expressions, and
function abstraction and application.  \footnote{Details about the
  language and its type system are
  provided in the supplemental material.}  To simplify the
presentation, in the following, we assume all variables have a single
unique binding-site.

$\lambda_{\S{qta}}$ types include standard base types like
\S{int, bool} etc., along with algebraic types like \S{lists} and
\S{trees} over these base types. Refinement types $\tau$, include
\emph{base refinements} and \emph{arrow refinements}. A base
refinement \{ $\nu$ : \S{t} $\mid \phi$ \} qualifies a term of base
type \S{t} with a refinement qualifier $\phi \in \Phi$. An arrow refinement
refines a function type, where the argument \S{x} can occur free in
the return type. Qualifiers ($\Phi$) is a set of first-order predicate
logic formulae over base-typed variables along with method predicates
($Q$), which are user-defined, uninterpreted function symbols such as
{\sf len} and {\sf ord} over lists used in our motivating example.
%   Uppercase variables like {\it P, Q, etc,.} range over method predicates. 
%   To precisely capture the semantics of higher-order and polymorphic terms, \calculus also contains \textit{Abstract Refinements} or \textit{Refinement Variables}, which are represented using later upper-case letters like $S, T$, etc. The language also has \textit{sorts} $s$, which gives the shape of the method-predicates or abstract refinement variables.
% %   Finally, \textit{Types} can be quantified with type-variables or abstract-refinements giving appropriate \textit{type schemas}.
A type context $\Gamma$ records term variables and library functions
$g$ with their types. It also records a set of propositions relevant
to a specific context.  

\section{Qualified Tree Automata}
\label{sec:qta}  

\begin{definition}[Positions in a term]
A position $p$ in a term $t$ is of the form $i.j.k....n$, a sequence
of positive integers describing a path from the root of $t$ to a
sub-term. This describes what symbols are present at each position,
relative to the root.
\end{definition}
\begin{wrapfigure}{r}{0.35\textwidth}
%\vspace*{.3in}
  %\ottmetavars\\[0pt]
\small \begin{tabular}{l l } % centered columns (4 columns)
%\emph{c} $\in$ \emph{Constants}  \\
%\textsf{x}, $p$ $\in$ \emph{Variables}  \\
%\textsf{Q} $\in Predicates_{T}$ \\
$p, p_i,... p_n$ $\in$ \emph{Position} \\
$\psi_a$ ::= & (\textit{Atoms})\\
\S{true} $\mid$ \S{false} & \\ 
$\mid$ $p$ = $p$  & (\textit{Syntactic equality}) \\
$\mid$ $p$ $\vDash$ $p$ & (\textit{Semantic entailment}) \\ 
$\psi \in \Psi$ ::= & \\
%\textsf{true} $\mid$ \textsf{false} $\mid$ \\
%$\Sigma_{T}$ $\mid$ \\
$\psi_a$ $\mid$ $\neg$ $\psi$ & \\
%$Q(\overline{p})$ $\mid$ $\psi$[p/p] \\
$\mid$ $\psi$ $\wedge$ $\psi$ $\mid$ $\psi \lor \psi$ & \\
$\sigma \in Schema$ ::= & \\
$\star$ $\mid$ $\star$ = $\star$ $\mid$ $\star$ $\vDash$ $\star$ & 
%$\mid$ $\psi$ $\Rightarrow$ $\psi$ \\
   % Note that we do not have \forall p. \phi or \forall p.
\end{tabular}%
\vspace*{-.1in}
\caption{QTA Constraints $\Psi$}
\label{fig:constraints}
\end{wrapfigure}

For easier comprehension, we give human-readable
labels to each number in a position, e.g., consider
Figure~\ref{fig:introduceqta} again. The sequences used in the
constraints like \S{arg.base = fun.in.base} are
positions. \S{type.ref} is a synonym for position \S{1.3},
\S{fun.in.base} for \S{2.1.2}, etc.\\

\begin{definition}[Constraints in \automata]
A QTA constraint $\psi \in \Psi$ is a predicate on terms in
\calculus. It is defined inductively over positions and Boolean
connectives, as shown in Figure~\ref{fig:constraints}.
A valid \textit{atomic constraint} $\psi_a$ includes Boolean constants
like \S{true} and \S{false}, as well as \textit{syntactic equality}
between positions, given by $p$ = $p$ and
\textit{semantic entailment} (\emph{Sem-ent}) over positions, given by
$p \vDash p$.  A \textit{constraint} $\psi$ is either a $\psi_a$ or a
constraint generated using Boolean connectives over other constraints.
Consequently, we can also classify \textit{atomic constraints} into
\textit{kinds} using three constraint schemas, $\star$,
$\star$ = $\star$ and $\star$ $\vDash$ $\star$, respectively.
\end{definition}

\begin{definition}[Qualified Tree Automata]
 A \textit{Qualified Tree Automata}, $\mathcal{A}$ defined over a finite ranked alphabet $\mathcal{F}$ derived from \calculus, is a tuple ($Q$, $\mathcal{F}$, $Q_f$, $\Delta$), where:
 \begin{itemize}
     \item $Q$ is a finite set of states.
     \item $Q_f \subseteq Q$ is a set of final states.
     \item $\Delta \subseteq Q^n \times \mathcal{F} \times \Psi \mapsto Q$, is a set of constrained \emph{transitions}. Each transition rule is of the form $f (q_1, q_2,...q_n) \T{\psi} q$,
  where $f \in \mathcal{F}$ with arity $n$, and a set of
  states $q_1, q_2, \ldots q_n \in Q$ and $\psi \in \Psi$ is a valid constraint. Here $q$ is the \textit{target state}.
 \end{itemize}
\end{definition}
\begin{wrapfigure}{l}{.40\textwidth}
  \small
 \centering % used for centering table
%\ottmetavars\\[0pt]
\begin{tabular}{l  l l} % centered columns (4 columns)
$\denotationNode{q}$ &::=  $\bigcup_{i}$ ($\denotationEdge{\delta_i}$) \\ &  where $\delta_i$ = \\
& ($f ({q_i}_1, {q_i}_2$,\ldots,${q_i}_n)$ $\T{\psi}$ $q$) \\
&  \\
$\denotationEdge{\delta}$  &::=  \{ $f$ 

$\overline{t_i}$ $\mid$  $t_i$ $\in \denotationNode{q_i}$,$f$ 

$\overline{t_i} \VDash \psi$, \\ 
& $i \in [1\ldots n]$ \} \\
& $\delta$ = ($f ({q}_1, {q}_2$,\ldots,${q}_n)$ $\T{\psi} q$) \\
$\denotationEdge{\delta_{\bot}}$ &  := $\varnothing$ \\
& \\
$\denotation{\mathcal{A}}$  ::= & $\bigcup_{i}$ \{$\denotationNode{q_i}$ $\mid$ $q_i \in Q_f$\}\\
\end{tabular}%
\caption{QTA denotation.}
\label{fig:denotation}
\end{wrapfigure}
\paragraph{\bf Language of a \automaton $\mathbb{L}$ ($\mathcal{A}$)}
The language accepted by a QTA $\mathcal{A}$, is the set of all terms
in \calculus with some successful run of $\mathcal{A}$. This is, in fact, the set of all well-typed \calculus terms, constructed using the
methods found in a library.  Formally, we define the language accepted
by $\mathcal{A}$ using its denotation $\denotation{\mathcal{A}}$ (see
Figure~\ref{fig:denotation}). The denotation of a state $q$,
$\denotationNode{q}$ is the set union of the denotations of each of
the transitions $\delta_i$, $\denotationEdge{\delta_i}$, for which $q$
is a target state. The denotation of a transition $\delta$
builds a set of all terms, using the symbol at the current transition
($f$) and terms in the denotation of states incoming in $\delta$, 
filtering all terms that do not satisfy the transition
constraint $\psi$. Symbols $f \in \mathcal{F}$ include all \calculus terms including variables (\S{x}), constants (\S{c}),  conditional expressions ( \S{if} b \ then \ e \ else \ e), function abstractions and applications, and 
types ($\tau$). 
Intuitively, the satisfaction of a constraint by a
term $t \VDash \psi$ maps syntactic equality constraints to equality
of symbols and semantic entailment between qualifiers to logical
entailment of FOL formulas.\NEW{We also have a special bottom transition $\delta_{\bot}$, whose denotation is an empty set}. Since a QTA can have multiple final
states given by the set $Q_f$, the language of a QTA is the union of
the denotation for all its \emph{final states}.

\section{Synthesis using \automata}
\label{sec:synthesis}

\subsection{Component-based synthesis using QTA}
%\NEW{\paragraph{Revised Synthesis Problem}}
\NEW{To formalize synthesis using QTAs, we revise our earlier definition.  First,
we define a consistency relation between a QTA $\mathcal{A}$ and a typing environment $\Gamma$.}

% \NEW{
% {\bf Transition at a position}:
% Given a  transition  $\delta$ = ($f ({q}_1, {q}_2$,\ldots,${q}_n)$ $\T{\psi} q$) in a QTA, and a position $p$ = $i.j...k$, we define a transition at the position $p$ from $q$, represented as ($q \blacktriangleright p$) inductively as follows: $q \blacktriangleright \epsilon$ = $\delta$ and $q \blacktriangleright i.j...k$ = $q_i \blacktriangleright j...k$ iff $i \in [1...n]$ else $q \blacktriangleright i.j...k$ = $\delta_{\bot}$.}
\NEW{\begin{definition}[Consistency between a QTA $\mathcal{A}$ and Type Environment $\Gamma$]
A type environment $\Gamma$ is \textit{consistent} with a QTA $\mathcal{A} = $ ($Q$, $\mathcal{F}$, $Q_f$, $\Delta$) iff $\forall e$ . $\Gamma (e)$~\footnote{Expression $e$ and variable binding $e$ are used interchangeably.}  = $\tau \iff 
\exists \mathcal{A}'$ such that $\mathcal{A}'$ is a sub-automaton of $\mathcal{A}$ and $e \in \denotation{\mathcal{A}'}$.
\end{definition}}
%We present a more operational definition of how %to constructs a $\Gamma$ consistent with an %\automaton in the supplemental material.}

%\begin{definition}
\NEW{
{\bf Revised Synthesis Problem}: Given a type environment $\Gamma$ that relates library functions $f_i = \lambda (\overline{x_{i,j}}). e_{f_i}$ with their refinement types
$f_i : \overline{(x_{i,j} : \tau_{i,j})} \rightarrow \{ \nu :  t_i \mid \phi_{i} \} \in \Gamma$, and a synthesis query  $\varphi = \overline{(y_i : \tau_i)} \rightarrow \{\nu : t \mid \phi \}$, 
a solution to a CBS problem is a QTA $\mathcal{A}$, such that forall all $e \in \denotation{\mathcal{A}}$, $\Gamma \vdash e : \varphi$ and $\Gamma$ is consistent with $\mathcal{A}$. }

\subsection{Synthesis Algorithm}
The main synthesis algorithm, {\sc QTASynthesize} is shown in
Algorithm~\ref{alg:enumerate}.  It takes as input an alphabet
$\mathcal{F}$, which includes  symbols from \calculus and a library of functions
with refinement type annotations; a synthesis query
specification $\varphi$, and a bound on the size of the terms
${k}$ to synthesize.  
% It also retutranspiled into
% judgment rules given in
% Figure~\ref{fig:typing} and Figure~\ref{fig:similarity}. 

The algorithm works in two phases: (1) an \textit{exploration} phase
adds states and transitions, expanding the automata. The resulting QTA
is then pruned/reduced by (2) a reduction phase.  The algorithm also
keeps track of \textit{similar} but not yet reduced transitions
through an equivalence set $\mathcal{E}$, lifting the subtyping
relation to an ordering relation between transitions in QTA.

\NEW{The output of the algorithm is a pair consisting of (i) a 
QTA, $\mathcal{A}_{\S{min}}$ for the synthesis query based on
$\mathcal{F}$ and the typing semantics of \calculus, such that the language of the QTA are solutions to the synthesis problem,
and (ii) a set of solution terms in \calculus, possibly using $\mathcal{F}$ that satisfies the query specification, 
The algorithm
returns a failure value ($\bot$) if it cannot find a solution within the
given max-depth $k$.}

\begin{wrapfigure}{r}{.5\textwidth}
  \centering
  \hspace*{.17in}
  \begin{algorithm*}[H]
    \small
%\begin{multicols}{2}
\SetAlgoNoLine
\SetNlSty{texttt}{(}{)}
\SetKwFunction{QTAsynthesize}{\textsc{QTASynthesize}}
\Indm\QTAsynthesize{$\langle \mathcal{F}, \varphi = \overline{(\mathsf{x_i} : \tau_i)} \rightarrow \{ \S{v} : \S{t} | \phi\}, k \rangle$}\\
\Indp
     \tcp{\textcolor{cyan}{Initialize}}
    \nl$\mathcal{A}_{0}$ $\leftarrow$ \textsc{WF} ($\mathcal{F}$, $\mathcal{A}_{\bot}$); $\mathcal{E} \leftarrow \varnothing$ \\
    %\nl$\overline{{q_0}_i}$ $\subseteq$ $Q_0$, |$q_0$| = \S{b} \\
    \tcp{\textcolor{cyan}{Check solution in Initial $\mathcal{A}_0$}}
    \nl\If { $\hat{Q} = $ {\sc NEmpty} ($ \mathcal{A}_0$)}{  
             \nl {\bf return} ($\mathcal{A}_0$, $\bigcup_{q \in \hat{Q}} \denotation{\mathcal{A}_q}$)
             %\nl {\bf return}  ($\mathcal{A}_0$, {\sc Terms} ($\mathcal{A}_0$))
    } 
    \tcp{\textcolor{cyan}{Iteratively explore-reduce-check}}       
    \nl {\bf return} \textsc{Enumerate} ($\mathcal{A}_0$, $\varphi$, $k$)\\[1mm]
\SetKwFunction{enumerate}{\textsc{Enumerate}}

\Indm\enumerate{$\langle \mathcal{F}, \mathcal{A}, \varphi = \overline{(\mathsf{x_i} : \tau_i)} \rightarrow \{ \S{v} : \S{t} | \phi\}, k \rangle$}\\
\Indp
    % \nl\While{true}{ 
        \nl \eIf{depth ($\mathcal{A}$) < k}    
        {  
           \nl $\mathcal{A}$ $\leftarrow$ \textsc{Transition} ($\mathcal{F}$, $\mathcal{A}$);\\ 
           \nl $\mathcal{A}$ $\leftarrow$ \textsc{Prune} ($\mathcal{A}$);\\ 
           \nl $\mathcal{E}$ $\leftarrow$ \textsc{Similarity} ($\mathcal{A}$, $\mathcal{E}$); \\ 
           \nl ($\mathcal{A}_{\S{min}}$, $\mathcal{E}$) $\leftarrow$ \textsc{Minimize} ($\mathcal{A}$, $\mathcal{E}$); \\ 
           %\nl \S{F} $\leftarrow$ \emph{Frontier}($\mathcal{A_{\S{min}}}$);\\  
           \nl\If { $\hat{Q} = $ {\sc NEmpty} ($\mathcal{A}_{\S{min}}$)}{  
             \nl {\bf return} ($\mathcal{A}_{\S{min}}$, $\bigcup_{q \in \hat{Q}} \denotation{\mathcal{A}_q}$)
             %\nl {\bf return} ($\mathcal{A}_{\S{min}}$, {\sc Terms} ($\mathcal{A}_{\S{min}}$))
             }   
           %\nl $\mathcal{A}$ $\leftarrow$ $\mathcal{A}_{\S{min}}$ \\
           \nl \textsc{Enumerate} ($\mathcal{A}_{\S{min}}$, $\varphi$, $k$)\\
        }{ %esle begin
            % \nl \If{$ (\hat{Q} = \S{SucessfulRun}$ ($\mathcal{A}, \varphi$)) $\neq \varnothing$}{  
            %  \nl {\bf return} $\bigcup_{q \in \hat{Q}} \denotation{\mathcal{A}_q}$
            % }{ 
                \nl {\bf return} $\bot$ \\[1mm]
            %}
         }%else-end
\SetKwFunction{nonemptyqta}{\textsc{NEmptyQTA}}

\Indm\nonemptyqta{$\langle \mathcal{A} \rangle$}\\
\Indp
    \nl $\hat{Q}$ $ \leftarrow \{ q_f  \mid q_f \in Q_f, \denotationNode{q_f} \neq \varnothing$ \} \\
    \nl {\bf return} $\hat{Q}$
    
    % }%while-end
    \caption{Main Synthesis Algorithm.}
\label{alg:enumerate}
\end{algorithm*}
\end{wrapfigure}

The algorithm begins (line 1) by initializing $\mathcal{E}$ to an
empty set and constructing an initial QTA covering all terms of size
one using a call to a well-formedness function ({\sc WF}), passing it
the library $\mathcal{F}$ and an empty automata $\mathcal{A}_{\bot}$.
This function is a deterministic implementation of the rules rules
shown in Figure~\ref{fig:typing}.  These rules formalize when to add
transitions corresponding to well-formed primitive types ({\sc
  wf-prim}), predicates ({\sc wf-pred}), and variables ({\sc wf-var}),
in the component library.  They also define how to add states and
transitions for each of the query arguments $\S{x}_i$ and their types
$\tau_i$. \NEW{Additionally, there is also a rule {\sc Q-goal} suggesting how to add a transition (and corresponding states) for the query specification. }
Consequently, initialization adds states and leaf
transitions for the arguments in the synthesis query $\varphi$,
library functions, base types, etc. 
Initialization also adds a final
state and transition with the top-level constraint corresponding to the
given query $\varphi$.  This generates the initial QTA
$\mathcal{A}_0$.

% \SJS{Do not introduce a new undefined functuin TERMS, rather explain the the terms are extracted based on the denoatational semantics. And later in Section implementation, we discuss the details of the extraction. We can keep the example as it is but not have a separate informally defined function {\sc Terms}.}
\NEW{Next, at line 2, the algorithm checks if the language of the initial QTA
$\mathcal{A}_0$ is non-empty using another routine {\sc NEmpty}
(lines 14 and 15), that collects all final states with a non-empty language using the earlier defined $\denotationNode{.}$ function.} 

% The \textit{skeleton} of a QTA is a FTA without the logical implication constraints. Intuitively, it checks if there is at least one term in the space of base-typed terms, where all refinement types are erased using an \textit{erasure} function~\cite{liquidextended}. If
% so, it extract all the terms in the language of the QTA using the term extraction function {\sc Terms}, defined below.} 
\NEW{If this set is non-empty, the algorithm extracts the set of solution terms in the language of $\mathcal{A}_0$ using the QTA denotation $\denotation{\ }$ (Figure~\ref{fig:denotation}), finally returning ($\mathcal{A}_0$, $\denotation{\mathcal{A}_0}$) (line 3).}
Otherwise (line 4), it calls procedure {\sc Enumerate} (lines 5-13)
that is the main \emph{explore-reduce-check} loop where the
construction (and reduction) of the QTA occurs. This procedure returns $\bot$, representing synthesis failure, if it is invoked when the max depth of the QTA has already been reached.

However, if the depth of $\mathcal{A}$ is less than the max-depth $k$,
{\sc enumerate} enters the exploration phase and expands $\mathcal{A}$
by adding new transitions, using a procedure {\sc Transition} (line
6). This is a deterministic implementation of the transition rules
in Figure~\ref{fig:typing} for capturing typing semantics for the
\calculus.  These new transitions update the QTA thus adding larger
terms.  

At this point, \textsc{Enumerate} does not further expand
$\mathcal{A}$; instead, the algorithm enters a \emph{reduction} phase.
It first reduces the size of $\mathcal{A}$, using the {\sc Prune}
function (line 7); we present its definition in
Section~\ref{sec:reductions}.  Next, the procedure performs
similarity checking and reduction using the {\sc Similarity}  (line 8) and
{\sc Minimize} (line 9) functions.  {\sc Similarity} is a
deterministic implementation of the rules given in
Figure ~\ref{fig:similarity}.   It returns an updated similarity set
$\mathcal{E}$ carrying all the similarity information in the
structures.  {\sc Minimize} (line 9) is an implementation of
the minimization inference rules ({\sc M-Trans}) and ({\sc M-QTA})
given in Figure ~\ref{fig:similarity}.  We describe these rules in the
next section.  In the process, the new QTA
may drop certain states and transitions that are logically equivalent
to more specific states or may merge several transitions.  The
result is a minimized automata $\mathcal{A_{\S{min}}}$.  Finally, 
\textsc{Enumerate} again checks (line 10) if the language of QTA
$\mathcal{A}_{\S{min}}$ is non-empty and returns the QTA along with the set of the extracted terms
in $\mathbb{L}$ ($\mathcal{A}_{\S{min}}$) (line 11) using the denotation 
$\denotation{.}$ $\mathcal{A}_{\S{min}}$.
Otherwise, the algorithm iterates again on
the minimized automata entering the \textit{exploration phase}.

\subsection{QTA Construction}
\begin{figure*}[htbp]
\begin{flushleft}
\bigskip
{\bf Well-formedness}\quad \fbox{\footnotesize
    $\begin{array}{c} 
    % A state in itself has no information now, so is always well-fomed
    \mathcal{F}, \mathcal{A} \vdash^{\S{wf}} f \in \mathcal{F} (\overline{q_i}) \T{\psi} q      
    \end{array}$
}
\end{flushleft}
\bigskip
\begin{minipage}{0.3\textwidth}
{\footnotesize
\begin{center}
\inference[{\sc wf-prim}]{\S{t} \in \S{T}_{\mathcal{F}} & q_{\S{t}} \notin Q}{\mathcal{F}, \mathcal{A} \vdash^{\S{wf}} \S{t} () \T{} q_{\S{t}}}
\end{center}
}
\end{minipage}
\begin{minipage}{0.3\textwidth}
{\footnotesize
\begin{center}
\inference[{\sc wf-pred}]{\phi \in \Phi_{\mathcal{F}} & q_{\phi} \notin Q}{\mathcal{F}, \mathcal{A} \vdash^{\S{wf}} \phi \T{} q_{\phi}}
\end{center}
}
\end{minipage}
\begin{minipage}{0.3\textwidth}
{\footnotesize
\begin{center}
\inference[{\sc wf-var}]{x \in \S{Vars}_{\mathcal{F}} & q_{x} \notin Q}{\mathcal{F}, \mathcal{A} \vdash^{\S{wf}} x \T{} q_{x}}
\end{center}
}
\end{minipage}
\bigskip
\bigskip
\begin{minipage}{0.4\textwidth}
{\footnotesize
\begin{center}
\inference[{\sc wf-base}]{(\tau \equiv \{ x : t \mid \phi \}) \in \tau_{\mathcal{F}} & q_{\tau} \notin Q} 
{\mathcal{F}, \mathcal{A} \vdash^{\S{wf}} \tau (q_x, q_t, q_{\phi}) \T{} q_{\tau}}
\end{center}
}
\end{minipage}
\begin{minipage}{0.4\textwidth}
{\footnotesize
\begin{center}
\inference[{\sc wf-arrow}]{(\tau_{\rightarrow} \equiv \tau_i \rightarrow \tau_j) \in \tau_{\mathcal{F}} \\
\tau_i, \tau_j \in \tau_{\mathcal{F}}  & q_{\tau_{\rightarrow}} \notin Q} 
{\mathcal{F}, \mathcal{A} \vdash^{\S{wf}} \tau_{\tau{\rightarrow}} (q_{\tau_i}, q_{\tau_j}) \T{} q_{\tau_{\rightarrow}}}
\end{center}
}
\end{minipage}
\begin{minipage}{0.4\textwidth}
{\footnotesize
\begin{center}
\inference[{\sc wf-t-abs}]{q_{\alpha}, q_{\tau} \in Q \\
\psi = q_{\S{tabs}} \blacktriangleright \S{tvar}.\S{type} = q_{\S{tabs}} \blacktriangleright \S{type.base}} 
{\mathcal{F}, \mathcal{A} \vdash^{\S{wf}} \S{tabs}(q_{\alpha}, q_{\tau}) \T{\psi} q_{\S{tabs}}}
\end{center}
}
\end{minipage}
\begin{minipage}{0.4\textwidth}
{\footnotesize
\begin{center}
\inference[{\sc Q-goal}]{\varphi = \overline{(\mathsf{x_i} : \tau_i)} \rightarrow \tau & q_{\S{term}_k}, q_{\tau} \in Q \\ q_{\S{goal}} \in Q_f \\
\psi = \textsc{SubType} (q_{\S{term}_k}\blacktriangleright{\S{type}}, q_{\S{goal}}\blacktriangleright{\S{type}})
}
{\mathcal{F}, \mathcal{A}  \vdash^{\S{wf}} \S{goal} (q_{\tau}, q_{\S{term}_k}) \T{\psi} q_{\S{goal}}}
\end{center}
}
\end{minipage}

\begin{flushleft}
{\bf Transitions}\quad \fbox{\footnotesize
    $\begin{array}{c} \mathcal{F}, \mathcal{A} \vdash  f \in \mathcal{F} (\overline{q_i}) \T{\psi} q \\ 
    % A state in itself has no information now, so is always well-fomed
    %\mathcal{A} \vdash^{\S{wf}} \textnormal{q}      
    \end{array}$
}
\end{flushleft}
\bigskip
% For each var/function in the environment, we have a transition with atleast one child, denoted by x.type
%var and constrant
\begin{minipage}{0.4\textwidth}
{\footnotesize
\begin{center}
\inference[{\sc e-var}]{ x : \tau \in \mathcal{F} & q_x \notin Q}
{\mathcal{F}, \mathcal{A} \vdash x (q_{\tau}) \T{} q_x}
\end{center}
}
\end{minipage}
\begin{minipage}{0.4\textwidth}
{\footnotesize
\begin{center}
\inference[{\sc e-const}]{ \vdash c : \tau \in \mathcal{F} & q_c \notin Q}
{\mathcal{F}, \mathcal{A} \vdash c (q_{\tau}) \T{} q_c}
\end{center}
}
\end{minipage}
\bigskip
\bigskip
\bigskip
\vspace*{.05in}
% Function and If
%Function call
\begin{minipage}{0.4\textwidth}
{\footnotesize
\begin{center}
\inference[{\sc e-app}]{q_f, q_a \in Q & \tau (\overline{q_i}) \T{} q_{\tau} \in \Delta\\
% \gamma, epsilon \vdash p1 <: p2
\psi = \textsc{SubType} (q_{f}\blacktriangleright{\S{out}}, q_{\S{app}}\blacktriangleright{\S{type}}) \wedge \\
\ \ \ \ \textsc{SubType} (q_{\S{a}}\blacktriangleright{\S{type}}, q_{f}\blacktriangleright{\S{in}})}
{ \mathcal{F}, \mathcal{A} \vdash \S{app} \ (q_{\tau}, q_f, q_a) \T{\psi} q_{\S{app}}}
\end{center}
}
\end{minipage}
% If
\begin{minipage}{0.4\textwidth}
{\footnotesize
\begin{center}
\inference[{\sc e-if}]{q_b, q_t, q_f \in Q & \tau (\overline{q_i}) \T{} q_{\tau} \in \Delta \\
\psi =  ((q_b \blacktriangleright {\S{ref}}) \wedge \textsc{SubType} (q_{t}\blacktriangleright {\S{type}}, q_{\S{if}}\blacktriangleright{\S{type}})) \wedge \\
\   \   \   \  (\neg (q_b \blacktriangleright{\S{ref}}) \wedge \textsc{SubType} (q_{f}\blacktriangleright{\S{type}}, q_{\S{if}}\blacktriangleright{\S{type}}))}{ \mathcal{F}, \mathcal{A} \vdash \S{if} \ (q_{\tau}, q_b, q_t, q_f) \T{\psi} q_{\S{if}}}
\end{center}
}
\end{minipage}
% Similarity and minimization 

\bigskip
{\footnotesize
\begin{equation}
\label{eq:sub}
\textsc{SubType} \textcolor{red}{(} \delta_i, \delta_j \textcolor{red}{)} = 
        \begin{cases}
            \textcolor{red}{(} \tau_i (\overline{q_i}) \T{\psi_i} q_{\tau_i}, 
            \tau_j (\overline{q_j}) \T{\psi_j} q_{\tau_j}\textcolor{red}{)} & \S{i.type.t} = \S{j.type.t}  \\& \wedge \ \S{i.type.ref} \vDash \S{j.type.ref} \\
            \textcolor{red}{(}\tau_{\rightarrow_i} (\overline{q_i}) \T{\psi_i} q_{\tau_{\rightarrow_i}}, 
            \tau_{\rightarrow_j} (\overline{q_j}) \T{\psi_j} q_{\tau_{\rightarrow_j}}\textcolor{red}{)} & \textsc{SubType} \textcolor{red}{(} \delta_j\blacktriangleright{\S{in}}, \delta_i\blacktriangleright{\S{in}}\textcolor{red}{)}   \\ & \wedge \ \textsc{SubType} \textcolor{red}{(} \delta_i\blacktriangleright{\S{out}}, \delta_j\blacktriangleright{\S{out}} \textcolor{red}{)} \\
            \textcolor{red}{(} \_, \_ \textcolor{red}{)} & \S{true}
        \end{cases}
\end{equation}
}
\caption{Selected rules for constructing transitions $\Delta$,  basis for {\sc WF} and {\sc Transition.}}
\label{fig:typing}
\end{figure*}

The transitions $\Delta$ for a QTA $\mathcal{A}$ are constructed
using the \textbf{Well-formedness} and \textbf{Transitions} judgments given in Figure~\ref{fig:typing}.
\NEW{The latter judgment holds if, given library $\mathcal{F}$ and  automata $\mathcal{A}$,  a new n-ary transition  can be added to $\mathcal{A}$  corresponding to an n-ary symbol $f \in \mathcal{F}$, with $q_1, q_2, ...q_n$ being the incoming states in the transition and $q$ being the target state, such that $\psi$ captures the typing constraints for the valid terms in the language of the transition $\denotationEdge{.}$}

\subsubsection{Well-formedness Rules}

Given the current alphabet $\mathcal{F}$, which contains the
component-library and the query $\varphi$, a current QTA
$\mathcal{A}$; and the well-formedness typing semantics in \calculus
(represented using $\vdash^{\S{wf}}$), judgments of the form
$\mathcal{F}, \mathcal{A} \vdash^{\S{wf}} f \in \mathcal{F}
(\overline{q_i}) \T{\psi} q$ direct how we can add a new transition
to $\mathcal{A}$. 
\NEW{These rules capture the conditions under which (leaf) transitions can be added to well-formed base types \S{t}, predicates $\phi$, variables, base and arrow refinement types, and polymorphic types. These rules closely follow the well-formedness typing rules for \calculus.}

% \NEW{Figure~\ref{fig:wf-example} shows these rules in action over a small example library of terms:
% \begin{tabbing}
% \noindent $\mathcal{F}$ =  \{\=\,\S{ f : (n : int)} $\rightarrow$ \S{(l : \{ v : [a]}\ |\  \S{len (v) > 0} \}) $\rightarrow$ \{ \S{v : [a]}\ |\ \S{len (v) > 0 }\} \\
% \>\S{g : (l : [a])} $\rightarrow$ \{ \S{v : [a]} \ |\ \S{len (v)} $\leq$ \S{len (l)} \} \\
% \>\S{xs : \{ v : [int]\ |\ len (v) > 0\} ; ys : [bool] \} }\}
% \end{tabbing}
% Let us assume that our query $\varphi$ is given by the refinement type:
% \begin{tabbing}
% \hspace*{.2in}(\S{xs} :  \{ v : [int] | len (v) > 0\}) 
% $\rightarrow$ \{$\nu$ : [int] | len ($\nu$) > len (xs)\}
% \end{tabbing}
% \noindent and the value of $k$, the term size is 2}.
% %g : (l : [char]) $\rightarrow$ [char]].}

\NEW{Rule {\sc wf-prim} adds transitions corresponding to each primitive
type \S{t} in the library.
For each such \S{t}, the rules creates a new state $q_{\S{t}}$ and adds a
nullary transition with symbol \S{t} and $q_{\S{t}}$ as the target
state. }
% This is shown in \textcircled{3} in Figure~\ref{fig:wf-example}, with all the $q_{\S{t}}$ merged int a single target node. 
Rule {\sc wf-pred} similarly adds leaf transitions for
refinement formula $\phi$ in some annotation in $\mathcal{F}$ or in the
query. 
% This can be seen in action in \textcircled{2} and 
% \textcircled{1} in a transition edge (\fbox{$
% \phi_{\S{f}}$} $\rightarrow \tau$) with $\phi_{\S{f}}$ as $\S{len (\nu) > 0}$.
Rule {\sc wf-var}, adds transition for variables in the library
or query. 

% an example is the transition and state corresponding to function \S{f} in Figure~\ref{fig:wf-example} (\textcircled{4}) with a transition (\fbox{\S{f}} $\rightarrow$ \textcircled{$q_f$}).}

Rule {\sc wf-base}, picks each well-formed base refinement type $\tau$
in the library or query annotation and creates a new state
($q_{\tau}$) for this type and adds a transition for $\tau$ with three
incoming states, corresponding to the three elements in a base
refinement \{ $x : t \mid \phi$ \}. Namely, $q_x$ for the
bound-variable $x$, $q_t$ for the base-type $t$ and $q_{\phi}$ for the
refinement formula $\phi$.  Note that the rule builds these states
constructed by earlier rules {\sc wf-var}, {\sc wf-prim} and {\sc
  wf-pred} as defined above.

% \NEW{Figure~\ref{fig:wf-example}, \textcircled{2} shows this rule in action for a base refinement type \{ $\nu$ : [t] | \S{len} ($\nu > 0$) \}}.
Rule {\sc wf-arrow} generates similar
transitions for each arrow refinement type, with a symbol
$\tau_{\rightarrow}$ and two incoming states for the argument-type and
the result-type for the arrow.
\NEW{Rule {\sc Q-goal} adds a special \S{goal} transition and a final state $q_{\S{goal}}$ for the given query and sub-automata rooted at $q_{\S{term}_{k}}$ for all terms of size $k$. The state $q_{\tau}$ captures the return type for $\varphi$. The transition constraint $\psi$ captures the standard subtype check between the type of the synthesized terms and the qoal's annotated return type, using a constrain generation function {\sc SubType}.}

\NEW{Our extended \calculus and typing semantics also include \textit{type} and \textit{refinement} abstractions, allowing parametric polymorphism in the refinement types setting~\cite{relref, liquidextended}. Consequently, the above rules also extend naturally to these abstractions and their corresponding type and refinement applications.
To illustrate, Rule {\sc WF-T-ABS} gives the rule to construct a transition representing a polymorphic type $\forall \alpha. \tau$, using the well-formedness semantics for type abstractions. Given two states, $q_{\alpha}$ for the type-variable and $q_{\tau}$ for the type with $\alpha$ occurring free, it constructs a transition representing the polymorphic type  $\forall \alpha. \tau$.\footnote{Please see the supplemental material for a complete description.}}

% \NEW{Figure~\ref{fig:wf-example}, \textcircled{1} shows this rule in action for the arrow refinement type for the example library function f : (n : int) $\rightarrow$ (l : \{ v : [a] | len (v) > 0 \}) $\rightarrow$ \{ v : [a] | len (v) > 0 \}.
% Finally, combined, these rules  generate the initial QTA $\mathcal{A}_0$ shown in ~\ref{fig:wf-example}(right).} 

% \begin{wrapfigure}{r}{.450\textwidth}
% \includegraphics[scale=.450]{fig/well-formed-example.png}
%  \caption{Application of {\sc Well-formedness} rules to construct initial QTA $\mathcal{A}_0$ for the example library.}
%  \label{fig:wf-example}
% \end{wrapfigure}
\subsubsection{Expression Transition Rules}
Transition judgments are similar in structure to the Well-formedness judgments, but simulate the refinement type judgments for the expressions in the \calculus. These rules say how to add transitions
corresponding to \calculus expressions along with their types. Each
n-ary expression transition has n+1 incoming
states, with the state at the position zero capturing the sub-automata
for the possible types of the expression.
% We use Figure~\ref{fig:e-example} to show some of these rules in action on the earlier example library

Rule {\sc E-var} and {\sc E-const}
adds transitions for variables and constants in the library along with
their types.  Note that {\sc E-var} rule will add transitions for both
scalars as well as variables bound to functions in the
library, consequently, the type $\tau$ in Rule {\sc E-var} will be modeled
either as a base refinement (via {\sc wf-base}) or an arrow refinement
(via {\sc wf-arrow}).

% \NEW{For instance, QTA sub-fragment labeled \textcircled{2} (see dashed circle) in Figure~\ref{fig:e-example} shows the application of E-var rule for the scalars \S{m}, \S{xs} and \S{ys} from the example library, creating transitions for each of these variables with associated types.}

Rule ({\sc E-app}) adds transitions related to function applications.
It assumes the states $q_f$ and $q_a$ are already present for
the function $f$ and argument $a$. It first constructs a transition
corresponding to the type ($\tau$) of the function-application using
the well-formedness judgments with $q_{\tau}$ as the target
state. The inferred transition uses {\sf app} as the symbol, with three
incoming states, corresponding to the resulting type ($q_{\tau}$),
function $q_f$ and argument $q_a$.  The point to note in this rule is
that constraints $\psi$ added to the transition, relating the three
sub-automata rooted at these states.  These constraints use an
auxiliary function {\sc SubType} (Equation~\ref{eq:sub}), which given
two transitions, returns constraints sufficient to capture the subtype
relation between their types.  For example, $\psi$ in
{\sc E-app} captures two main relations.  The first is a subtyping constraint
between a function's formal argument and the actual expression passed as
the argument ($q_{\S{a}}\blacktriangleright{\S{type}},
q_{f}\blacktriangleright{\S{in}}$).  This specifies that
given a state $q_{\S{a}}$,
($q_{\S{a}}\blacktriangleright{\S{type}}$) represents the transition
at position {\bf type} from $q_{\S{a}}$.  The second subtype relation is between
the function's result type and the type of the expression
($q_{f}\blacktriangleright{\S{out}},q_{\S{app}}\blacktriangleright{\S{type}}$).  
% \begin{wrapfigure}{r}{.450\textwidth}
% \includegraphics[scale=.45]{fig/e-example.png}
%  \caption{Application of {\sc Transition rules} to construct larger QTA from $\mathcal{A}_0$}
%  \label{fig:e-example}
% \end{wrapfigure}
The details of the auxiliary
operation {\sc SubType} are given in Equation~\ref{eq:sub}.  Note that
the {\sc E-App} rule is sufficiently general to allow synthesis to
support \textit{partial} (i.e., higher-order) function applications.
% \NEW{The QTA fragments labeled \textcircled{3} in Figure~\ref{fig:e-example} shows this rule in action over the function $\S{f}$ from the initial QTA $\mathcal{A}_0$ (\textcircled{1}) and the automata rooted at state $q_a$ for the scalars (\textcircled{3}). Transition constraints shown with label {\sc SubType} capture the required subtyping constraints between \S{f}'s input types and the scalar arguments (\S{arg.base} = \S{fun}.$\S{in}_1$.\S{base} and \S{arg.ref} $\vDash$ \S{fun}.$\S{in}_1$.\S{ref}). It also captures the subtyping constraints between the \S{app}'s return type and the function's output type (\S{fun.out.base} = \S{type.base} and \S{fun.out.ref} $\vDash$ \S{type.ref}).}

% \NEW{Figure~\ref{fig:e-example} shows how application of these rules constructs larger QTA from $\mathcal{A}_0$, depicting the expansion phase of the the synthesis algorithm.}

Rule {\sc E-if} builds a transition for a conditional expression using
the sub-automata for the Boolean condition (rooted at state $q_b$),
and the true and the false branches, rooted resp. at $q_t$ and
$q_f$. The constraints on the transition captures the standard
refinement typing semantics for conditional expressions. The true
branch adds the constraints that the Boolean condition is true while
the false branch specifies the negation.~\footnote{The complete set of typing judgments, construction rules and example showing the QTA construction using these rules are provided in the supplemental material.}

\section{QTA Reductions}
\label{sec:reductions}

\subsection{\textsc{Prune}}

The QTA formulation in Section~\ref{sec:pftree} accepts only
well-typed terms from the \calculus. However, we can make the
synthesis procedure more efficient by eagerly reducing portions of the
automata (i.e. sub-automata) which are irrelevant to the construction
of any solution.

\begin{figure*}[t]
\begin{flushleft}

{\bf Pruning}\quad\fbox{\footnotesize
     $\begin{array}{c} 
        \mathcal{A} \vdash \Delta \rightsquigarrow \Delta' \ \ \ \  \mid \ \ \ \ \mathcal{A} \vdash  \delta \rightsquigarrow^{\psi_a} \delta' 
      \end{array}$
           
} 

\end{flushleft}
\bigskip
{\footnotesize
\begin{minipage}{0.40\textwidth}
\begin{center}
\inference[{\sc p-trans}]{ \delta \in \Delta \equiv f_i (q_{i1}, q_{i2},...q_{in}) \T{\psi} q \\ 
\psi \equiv \bigwedge_{j \in [1\ldots m]} \psi_i \\ 
\mathcal{A} \vdash  \delta \rightsquigarrow^{\psi_j} \delta_{\S{r}}
} 
%\delta' = f (q_1, q_2, \ldots q_j \ldots q_n) \T{} q_j} 
{\mathcal{A} \vdash \Delta \rightsquigarrow \Delta[\delta_{\S{r}}/\delta]} 
\end{center}
\end{minipage}
\bigskip
\begin{minipage}{0.40\textwidth}
\begin{center}
\inference{ \psi_j \equiv p_1 = p_2 \\ 
\delta_{\S{r}} = \sqcap_{\S{Syntax}} (\delta\blacktriangleright p_1, \delta\blacktriangleright p_2) } 
 { \mathcal{A} \vdash  \delta \rightsquigarrow^{\psi_j} \delta_{\S{r}}}[{\footnotesize{\sc p-syn-eq}}]   
\end{center} 
\end{minipage}

\begin{minipage}{0.40\textwidth}
\begin{center}
\inference{ \psi_j \equiv p_1 \vDash p_2 \\ 
\delta_{\S{r}} = \sqcap_{\S{Semantics}} (\delta\blacktriangleright p_1, \delta\blacktriangleright p_2) } 
 { \mathcal{A} \vdash  \delta \rightsquigarrow^{\psi_j} \delta_{\S{r}}}[{\footnotesize{\sc p-sem-ent}}]   
\end{center} 
\end{minipage}

}

\begin{flushleft}
{\bf Similarity}\quad \fbox{\footnotesize
  $\begin{array}{c} 
\mathcal{A} \vdash^{\S{\bf sim}} \delta_i \lesssim \delta_j \ \ \ \  \mid \ \ \ \ \mathcal{A} \vdash \mathcal{E} \rightsquigarrow \mathcal{E}'
\end{array}$
}
\end{flushleft}
\bigskip
\begin{minipage}{0.4\textwidth}
{\footnotesize
 \inference[\footnotesize {\sc s-trans}]{ \psi_{<:} = \textsc{SubType} \textcolor{red}{(} \delta_i\blacktriangleright \S{type}, \delta_j\blacktriangleright \S{type} \textcolor{red}{)} \\  \delta_i\blacktriangleright \S{type} \sqcap_{\psi_{<:}} \delta_j\blacktriangleright \S{type} \neq \delta_{\bot}}
			  { \mathcal{A} \vdash^{\S{{\bf sim}}} \delta_i \lesssim \delta_j} 
}
\end{minipage}
\begin{minipage}{0.40\textwidth}
{\footnotesize
\inference[{\sc s-eq}]{ (\delta_i, \delta_j) \notin \mathcal{E} \\
\mathcal{A} \vdash^{\mathbf{sim}} \delta_i \lesssim \delta_j } {\mathcal{A} \vdash \mathcal{E} \rightsquigarrow \mathcal{E} \cup \{(\delta_i, \delta_j)\}}
}
\end{minipage}

\bigskip

\begin{flushleft}

{\bf Minimization}\quad\fbox{\footnotesize
     $\begin{array}{c} 
        (\mathcal{A}, \mathcal{E})  \vdash \Delta \rightsquigarrow \Delta' \ \ \ \  \mid \ \ \ \ \vdash (\mathcal{A}, \mathcal{E})   \rightsquigarrow (\mathcal{A}', \mathcal{E}') 
      \end{array}$
           
} 

\end{flushleft}
\bigskip
{\footnotesize
\begin{minipage}{0.40\textwidth}
\begin{center}
\inference[{\sc m-trans}]{\delta_i, \delta_j \in \Delta & (\delta_i , \delta_j) \in \mathcal{E} \\
\delta_i \equiv f (q_1, q_2, \ldots q_j \ldots q_n) \T{\psi_i} q_i  \\
\delta_j \equiv f' (q_1', q_2', \ldots q_m') \T{\psi_j} q_j  \\
%c_i \in \mathcal{C}(Q, \mathcal{E}), j \in [1..n]. q_j \in c_i & q_j \neq \underbar{$c_i$} \\
\Delta' = \bigcup_k \textcolor{red}{\{} \delta_k[q_j \mapsto q_i] \\ 
\ \ \ \ \ \ \ \ \ \ \ \  \  \   \    \ \mid \delta_k = \hat{f} (\hat{q_1}, {\bf q_j}, \ldots \hat{q_m} ) \hookrightarrow \hat{q_k}\textcolor{red}{\}} } 
%\delta' = f (q_1, q_2, \ldots q_j \ldots q_n) \T{} q_j} 
{(\mathcal{A}, \mathcal{E}) \vdash \Delta \rightsquigarrow (\Delta \cup \Delta') \setminus 
\{\delta_j\}} 
\end{center}
\end{minipage}
\bigskip
\begin{minipage}{0.40\textwidth}
\begin{center}
\inference{ {\footnotesize \mathcal{A} \equiv (Q, \mathcal{F}, Q_f, \Delta}) & {\footnotesize  (\mathcal{A}, \mathcal{E}) \vdash \Delta \rightsquigarrow^{\star} \Delta'}} 
 { \vdash (\mathcal{A}, \mathcal{E}) \rightsquigarrow ((Q, \mathcal{F}, Q_f, \Delta'), \varnothing)}[{\footnotesize{\sc m-qta}}]   
\end{center} 
\end{minipage}
}
\caption{Selective rules Similarity inference and QTA Minimization.}
\label{fig:similarity}
\end{figure*}

The inference rules for pruning irrelevant code, which form the basis
for the {\sc Prune} routine in Algorithm~\ref{alg:enumerate}, are given
as judgments in Figure~\ref{fig:similarity}. These rules
have two judgment forms, $\mathcal{A} \vdash \Delta \rightsquigarrow
\Delta'$ takes the current transition set and reduces it to a pruned
set. The other judgment form, $\mathcal{A} \vdash \delta
\rightsquigarrow^{\psi_a} \delta'$ reduces an individual transition by
an atomic constraint $\psi_a$ giving a pruned transition.

The {\sc p-trans} rule takes a transition $\delta$ with transition
constraint $\psi$. The rule assumes $\psi$ as a conjunction of atomic
\NEW{constraints} $\psi_j$ and reduces the chosen transition $\delta$ by
each $\psi_j$ (defined next) and updates the original $\delta \in
\Delta$ with the reduced $\delta_{\S{r}}$.

Reduction of a transition by an atomic constraint is defined in the
remaining two rules, matching the shape of the atomic constraint. Rule
{\sc p-syn-eq} handles syntactic equality constraints over positions
$p_1$ and $p_2$. It performs a syntactic intersection~\cite{ecta, tata} over the two
transitions at these position.  Syntactic intersection
($\sqcap_{\S{Syntax}}$) is a standard tree intersection
operation. Intuitively, it compares the two transitions for syntactic
equality of transition symbols while recursively \NEW{intersecting each incoming state in the transition.}

Rule {\sc p-sym-ent} handles the alternate case, when $\psi_j$
expresses a semantic entailment. When the constraint is $p_1 \vDash
p_2$, standard equality based intersection is ineffective, as the
formulas at positions $p_1$ and $p_2$ cannot be compared
syntactically. Thus, we define a semantic intersection operation
$\sqcap_{\S{Semantics}}$, which only compares transitions having
refinement qualifiers and compares them logically, checking the
logical entailment of the formula at position $p_2$ by the formula at
position $p_1$, keeping the transition at lower position
$p_1$. Intuitively, this operation compares transitions modeling
refinements of two types, and keeps the sub-type transition if the logical constraint hold, i.e. 
$\delta_{\S{r}} = \delta \blacktriangleright p_1$, 
if the formula at $\delta \blacktriangleright p_2$ implies \NEW{the} formula at $\delta \blacktriangleright p_1$. Otherwise, it returns a special bottom transition, i.e. 
$\delta_{\S{r}} = \delta_{\bot}$. The operation also runs a normalization routine and
trims all $\bot$ transitions.  Combined, these rules return a $\Delta$
with transitions reduced to $\delta_{\bot}$, forming the basis of the {\sc
  Prune} procedure in {\sc QTASynthesize}.

\subsection{\textsc{Similarity} and \textsc{Minimize}}
\label{subsec:similarity}

The similarity inference rules are given in the {\sc Similarity} judgments in Figure~\ref{fig:similarity}. The {\sc S-trans} rules suggests when two transitions are \textit{similar} based on their associated type sub-automata. 
% Formally, we define \textit{Similarity} as follows:
% \begin{definition}[Similarity between transitions]
% \label{def:sim}
% Given a \automaton $\mathcal{A}$ and two transitions $\delta_i =  f_i (q_{\tau}, \overline{q})$ \T{\psi_i} $q_i$ and $\delta_j =  f_j (q_{\tau}, \overline{q})$ \T{\psi_j}  $q_j$. 
% $\delta_i$ is similar to $\delta_j$, denoted by the relation $\delta_i \lesssim \delta_j$ iff $\delta_i \sqcap_{\psi_{<:}} \delta_j$ $\neq \delta_{\bot}$. 

The crux of the definition rests on the constraint $\psi_{<:}$ =
($\textsc{SubType} (\delta_i\blacktriangleright{\S{type}},
\delta_j\blacktriangleright{\S{type}}$).  This constraint checks if
the transition at position \S{\bf type} for $\delta_i$ and \S{\bf
  type} for $\delta_j$ are related by the standard subtype checks in a
\calculus typing semantics.  Rule ({\sc S-eq}) picks transition pairs
from $\mathcal{A}$ not in $\mathcal{E}$ and checks their similarity,
adding the pair to the similarity set $\mathcal{E}$ if the similarity
relation holds.

% Figure~\ref{fig:refmotivate} shows such a similar set of transitions using green partial arrow $\rightharpoonup$ in our QTA for the motivating example in Figure~\ref{fig:refmotivate}. 

% \begin{figure}
% \includegraphics[scale=.45]{fig/similar.png}
%  \caption{\TODO{A simple figure will also work}Transition Similarity}
%  \label{fig:similar}
% \end{figure}
\emph{Similarity Reduction}: The above definition of similarity
between transitions gives us an algorithmic way to minimize a QTA,
reducing all similar transitions but one.  The \textbf{Minimization}
rules in Figure~\ref{fig:similarity} define how to minimize a given
QTA using the similarity relation between transitions.  The rules have
two judgment forms; $(\mathcal{A},\mathcal{E}) \vdash \Delta
\rightsquigarrow \Delta'$, which takes the original QTA $\mathcal{A}$
along with the similarity relation $\mathcal{E}$ and translates the
original transition set to a new one, 
dropping the transition(s) with the super-type while keeping the 
subtype transitions; this is
represented in Rule {\sc M-trans}. It also updates the transition set
$\Delta$ such that whenever the state $q_j$ from the supertype
transition, $\delta_j$  flows in originally, in $\mathcal{A}$,
the rule now adds and incoming edge from $q_i$.

The second judgment defines how the complete automata is minimized and
the similarity set is updated on minimization. This is shown in the
{\sc M-qta} rule, which creates a minimized transition set $\Delta'$
using the transitive closure of the transition update rules ($\Delta
\hookrightarrow \Delta'$) defined by ($\hookrightarrow^{\star}$). The
updated QTA symbol set $\mathcal{F}$ remains the same as the
original. Finally, the updated similarity set becomes empty since all
similar states are either deleted or merged with other states.

% To illustrate, consider Figure~\ref{fig:reductions} again, using {\sc
%   s-trans} rules between the transitions for \S{x} and \S{y}, shown in
% the box with label {\bf Similar}, the $\psi_{<:}$ constraint is shown
% in the blue box. Since, the constraint holds (under variable renaming),
% these two transitions are marked as similar (shown by the
% green arrow) and added to $\mathcal{E}$. The {\sc m-trans} rule finally
% uses this similarity information to remove the transition for \S{y}
% while keeping \S{x}.

Note that the equivalence of terms (and hence sub-automata) is a
stronger property than similarity.  Consequently, 
similarity inference and reduction also reduces any
equivalent sub-automata, thus allowing us to prune away both
kind of redundant terms (see Figure~\ref{fig:explosion}) while
giving an efficient enumeration in a reduced search space.

%\AM{Needs redefinition , which will be easier to prove.}
% \begin{theorem}[Completeness Similarity Reduction]
%    The Similarity Reduction does not violate the completeness of the original automata modulo the synthesis query. Formally, given a refinement typed library $\mathcal{F}$, and a query $\varphi$, and given a QTA $\mathcal{A}$ constructed using {\sc WF} and {\sc Transition} rules. If {\sc Minimize} ($\mathcal{A}$, {\sc Similarity} ($\mathcal{A}$)) = $\mathcal{A}_{r}$ and $\nexists e \in \denotation{\mathcal{A}_r}$, such that $\mathcal{F}, \mathcal{A} \vdash e : \varphi$ then $\nexists e \in \denotation{\mathcal{A}}$, such that $\mathcal{F}, \mathcal{A} \vdash e : \varphi$.

%    % if $\exists e \in \mathbb{L} (\mathcal{A})$, such that $\Gamma \vdash e : \phi$, and, the Similarity reduction for $\mathcal{A}$ gives $\mathcal{A}_{red}$, then $\exists e' \in \mathbb{L} (\mathcal{A}_{red})$, and  $\Gamma' \vdash e' : \phi$.
% \end{theorem}

\section{Soundness and Completeness}
\label{sec:proofs}
For a given upper bound {\it k} on the size of programs being
synthesized, the {\sc QTASynthesize} algorithm is both sound and
complete assuming the
validity of each library function against their
specifications.\footnote{Proofs can be found in the supplemental material.}

% \SJ{These two definitions need to be changed to conform to the new declarative definition of the synthesis problem}
% \paragraph{\bf Soundness}
% Programs synthesized by the {\sc QTASynthesize} procedure are correct with
% respect to the provided query specification $\varphi$ assuming the
% validity of each library function against their
% specifications.\footnote{Proofs can be found in the supplemental material.

%\TODO{Supplemental: Rework the proofs for these new statements.}
\begin{theorem}[Soundness]
Given a type environment $\Gamma$ that relates library functions $f_i = \lambda (\overline{x_{i,j}}). e_{f_i}$ with their refinement types
$f_i : \overline{(x_{i,j} : \tau_{i,j})} \rightarrow \{ \nu :  t_i \mid \phi_{i} \} \in \Gamma$, and a synthesis query  $\varphi = \overline{(y_i : \tau_i)} \rightarrow \{\nu : t \mid \phi \}$, 
if {\sc QTASynthesize} ($\Gamma, \varphi, \S{k}$) =($\mathcal{A}_{\S{min}}$, \S{Terms} = $\mathrm{\{ }$e $\mid$ e $\in$ $\denotation{\mathcal{A}_{\S{min}}}$ $\mathrm{\}}$), then $\forall e \in \S{Terms}, \Gamma \vdash e: \varphi$, where $\Gamma$ is consistent with $\mathcal{A}_{\S{min}}$. 
\end{theorem}

%The {\it completeness} and {\it optimality} arguments are inter-connected. 
% Our completeness theorem ensures that if there exists a non-empty set
% of valid \calculus terms containing fewer than \S{k}+1
% library function calls that satisfy $\varphi$, then {\sc QTASynthesize} will
% terminate and produce some term from this set.
\begin{theorem}[Completeness]
\label{thm:completeness}
    Given a type environment $\Gamma$ that relates library functions $f_i = \lambda (\overline{x_{i,j}}). e_{f_i}$ with their refinement types
$f_i : \overline{(x_{i,j} : \tau_{i,j})} \rightarrow \{ \nu :  t_i \mid \phi_{i} \} \in \Gamma$, and a synthesis query  $\varphi = \overline{(y_i : \tau_i)} \rightarrow \{\nu : t \mid \phi \}$, 
if {\sc QTASynthesize} ($\Gamma, \varphi, \S{k}$), = $\bot$, then $\nexists$
    a term $e \in \denotation{\mathcal{A}_{\S{complete}}}$ containing fewer than $k+1$ library function calls, such that
    $\Gamma \vdash$ $e$ : $\varphi$ and $\Gamma$ is consistent with $\mathcal{A}_{\S{complete}}$. Where $\mathcal{A}_{\S{complete}}$ is the complete QTA of size $k$, for the given $\Gamma$, generated without any reduction.
    % such that   $\Gamma \vdash$ $e$ : $\varphi$. \SJ{Don't we need to parameterize $\vdash$ with $\matcal{A}$?}
\end{theorem}

\section{Synthesis Details and Implementation}
\NEW{In the QTA construction and pruning rules and the {\sc QTASynthesize} algorithm above, for ease of illustration, we have abstracted away several details about how we maintain variable scoping, infer types for transitions, and do efficient term extraction. Below, we discuss some of these in detail. For illustration, we will consider a library $\mathcal{F}$ = [\S{f} : (n : int) $\rightarrow$ (l : \{ v : [a] | len (v) > 0 \}) $\rightarrow$ \{ v : [a] | len(v) = len(l) \};
xs : \{ v : [int] | len (v) = 1\} ; ys : [char]; 
g : (l : [char]) $\rightarrow$ [char]].
Figure~\ref{fig:terms} shows a portion of minimized QTA for terms of size two with transition \fbox{\S{app}} for this library.
}

\subsection{Variable Scoping and Typing Environment in QTA}
\NEW{The details of variable scoping during QTA construction and pruning are important to understand how enumeration works with refinement types in a QTA. 
To simplify scoping decisions and manage the typing environment, we made several design choices.
First, we require terms in our synthesis language \calculus to be in A-normal form, and to have a unique binding variable $\S{t}_i$ for each application and conditional term. 
We also refactor each library function specification, alpha-renaming all bounded variables in the arguments with unique argument(s). This allows us to avoid unwanted variable capture across library functions without explicitly keeping track of scope information about bound and free variables. }

\begin{wrapfigure}{r}{.40\textwidth}
\includegraphics[scale=.450]{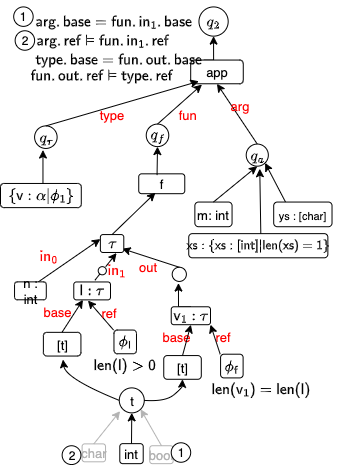}
\caption{ Partial QTA for the example library, variables renamed to avoid variable capture.} 
\label{fig:terms}
\end{wrapfigure}

\NEW{Additionally, to build the typing environment, each term binding variable in ANF  is ascribed a set of possible types that can be associated with it. To implement this structurally in QTA, we extend each n-ary expression transition (e.g.\S{app} transition) in the QTA, to (n+1) arity with an additional incoming edge for the \textcolor{red}{type} of the resulting expression (e.g., function application term). See, for example, the arrows with label \textcolor{red}{type} in Figure~\ref{fig:terms} for the \S{app} transition that has an edge ($q_{\tau} \rightarrow$ \fbox{\S{app}}),
$q_{\tau}$ to represent a set of valid types that can be ascribed to the application term (described next). 
Finally, we build a global typing environment mapping each expression type (annotated and inferred) with the binding variable using a typing environment construction function. This function is derived from a relation relating QTAs to typing environments; details are provided in the supplemental material. }
\subsection{Resolving Refinement Predicates for \textcolor{red}{type} Edges in Transitions}
\NEW{QTA construction and pruning also rely on inferring a set of feasible types for each transition. For instance see the incoming state ($q_{\tau} \rightarrow$ \fbox{\S{app}}) in the example. The possible set of types is shown using a transition (\fbox{$\{ \nu : \alpha \mid \phi_1\}$}
$\rightarrow$ \textcircled{$q_{\tau}$}). }

\NEW{Inferring this type set precisely requires inferring the base type (shown by a type variable $\alpha$ in the example) and the refinement predicate (here $\phi_1$). The former is straightforward, given the transition constraints, using 
syntactic comparison between terms at constrained location, e.g., the given constraint relating function and argument type structure allows us to infer in this case that the type of the application term is [$t$], a list over some base type $t$. 
Inferring the refinement predicates is more convoluted and needs some elucidation.
We assign abstract predicates to each missing refinement, e.g. $\phi_1$ in the example, for the refinement predicate. 
The logical constraint over transitions allows us to maintain logical \textit{implication} relations over these abstract predicates with other similarly constrained predicates.}

\NEW{For instance,  when inferring the type for the term \S{let} $\mathsf{t}_1$ = \S{f \ m  \ x} (under a typing environment constructed over the refactored QTA as explained earlier), we have two \textit{implication} constraints (i) \S{len} ($\S{v}_1$) = \S{len l} $\implies \phi_1$  and (ii) \S{len (xs) = 1} $\implies$ true. When solving for $\phi_1$, in the context, we can unify \S{l} with \S{xs}, and  with $\S{v}_1$ with $t_1$ to precisely ascribe/infer $\phi_1$ as \S{len} ($\S{t}_1$) = \S{len (xs)}  $\wedge$ \S{len (xs)} = 1 which reduces to \S{len} ($\S{t}_1$) = 1. This satisfies both constraints (i) and (ii). We can generally have a set of such ascribed types for cases with multiple functions and arguments in an \S{app} transition.}

% Note that when constructing the next transition corresponding to programs of size two (like \S{g \ ( f \ m \ x)}), we use this set of ascribed type to prune out  ill-formed terms like \S{g \ ( f \ m \ x)}) using the {\sc p-trans} rule.

% For example, when inferring the type for the application term \S{let} $t_1 = \S{app} \ (\S{f} \ \S{xs})$ (under a typing environment constructed over the refactored QTA as explained earlier), using the {\sc E-app} rule we generate two implication relations: (i) $\forall \nu$. \S{len} ($\nu$) > 0  $\implies$ \S{len} ($\nu$) > 0 and (ii) \S{len} ($t_1$) > 0$\implies$ $\phi_1$. 
% When solving for these relations, we find the strongest $\phi_1$ satisfying these constraints while also utilizing the fact that these predicates have to be generated from the refinement predicates in the type for one of the functions in the language of the incoming state $q_f$ (please see rule E-app in Figure~\ref{fig:typing}), via substitution of one of the arguments in the language of the state $q_a$. For instance here
% (i) is trivially true, when solving for $\phi_1$ using (ii), in the context, we can precisely ascribe/infer $\phi$ as \S{len} ($t_1$) > 0. We can generally have a set of such ascribed types for cases with multiple functions and arguments in an \S{app} transition.

\subsection{QTA Term Extraction}

\NEW{
Once a minimized QTA $\mathcal{A}_{\mathsf{min}}$ is constructed for a given size k, library $\mathcal{F}$, and the given $\phi$, Algorithm {\sc QTASynthesize} checks for the non-emptiness of the QTA and if successful, returns the QTA and the terms in the language of the QTA. 
The QTA denotation function call, in the Algorithm, is defined in Figure~\ref{fig:denotation} and gives a naive term extraction strategy where: a) we enumerate all terms based on underlying unconstrained tree-automata, b) filter out those terms which violate the constraints of each transition.
Unfortunately, even on minimized QTA, this will face a challenge in scaling to interesting scenarios. To address this, our implementation uses a more clever strategy employing a hierarchical combination of; a) a lazy choice of concrete types for polymorphic type variables, based on equality constraints~\cite{ecta}. This allows an efficient enumeration of valid base-typed terms. However, these well base-typed terms may still contain ill-typed terms in the refinement-type world. To filter out these terms: b) we have a secondary enumeration (using the denotation function) over these valid, based-typed terms, discarding terms that do not satisfy the logical constraints.}

% \TODO{We need to expand this using an example}
% \AM{We now have the figure and explanation in hard format bring it here next}
% To illustrate, consider a library $\mathcal{F}$ = [\S{f} : (n : int) $\rightarrow$ (l : \{ v : [a] | len (v) > 0 \}) $\rightarrow$ \{ v : [a] | len (v) > 0 \}; xs : \{ v : [int] | len (v) > 0\} ; ys : [bool]; g : (l : [char]) $\rightarrow$ [char]].
\NEW{Figure~\ref{fig:terms} shows \fbox{\S{app}} containing both equality (\textcircled{1}) and logical constraints (\textcircled{2}). While enumerating monomorphic type for \S{f}, our strategy first uses \textcircled{1} to deduce that that the type variable \S{t} cannot reduce to the base type \S{bool} as there are no argument under \textcircled{$q_a$} of \S{[bool]}. This is shown by a greyed out transition 
(\fbox{\S{bool}} $\rightarrow$ \textcircled{t}).
We next enumerate the remaining terms using the $\denotation{.}$ function, this further filters out the choice for \S{t} to be \S{char}, as type for \S{ys} does not satisfy \textcircled{2}, shown by a corresponding greyed out transition (\fbox{\S{char}} $\rightarrow$ \textcircled{t}). This leaves only a single enumeration choice for t as \S{int}, giving a unique monomorphic type for f.}

% \paragraph{Termination of {\sc QTASynthesize}}

% \NEW{Note that {\sc QTASynthesize} algorithm (Algorithm~\ref{alg:enumerate} is guaranteed to terminate in
% a finite number of steps for a given bound k. For any given max-depth k, and a starting minimized automaton of size m. The {\sc Transition} routine potentially increments the depth to m' >= m.
% Although, {\sc Prune} and {\sc  Minimize} can potentially collapse states and transitions, and hence indeed, can potentially reduce the depth to a value less than m.
% However, {\sc Transition},  {\sc Prune}, and {\sc Minimize} are deterministic implementations of the rules in Figures ~\ref{fig:typing} and ~\ref{fig:typing2}. The implementation of these functions memoize previously pruned transitions to avoid infinite cycles among these three procedures. Hence, since there are finitely many library functions and the algorithm is bounded by a finite maximum size k, we are guaranteed that Algorithm always terminates.
% }

\subsection{Implementation}
\label{sec:impl}

We have implemented these ideas in a tool (\name) that comprises
approximately 5KLoC in OCaml.  The input to \name\ is a specification
file containing a library of functions and data constructors, along
with their specifications, followed by a goal query.  The
specification and query languages are type-based, using refinement
types for pure functions~\cite{JV21}, extended with support for
polymorphic type specifications.  We rely on OCaml lexing and parsing
libraries OCamllex~\cite{ocaml} for handling the front end of our
query specification language. We use Z3~\cite{z3} to discharge SMT
queries. 

\paragraph{\name{} Library Specifications}
\name\ performs component-based synthesis using libraries annotated
with refinement-type specifications.  Fortunately, there are multiple
open-source projects that provide such specifications for our
use~\cite{vocal, JV21, ecta, cobalt-tech}.  Our experiments adapt
approximately 300 refinement type-annotated library functions drawn
from these projects.  These functions span operations on data
structures (e.g., arrays, lists, trees, queues, vectors,
zippers, byte strings, etc). We additionally include approximately 40
more specialized functions that target a specific database application
class ~\cite{database-examples}.  Approximately, 25\% of these library
functions are higher-order.

% \section{Implementation and Evaluation}
% \label{sec:eval}
% \AM{We already, have a finite/small representation of the infinite/large set of concrete programs given by the Typechecking rules.
% Till now these rules are applied at the meta level, from the programs, however in this work we make these rules a finite representation of the programs, using QTA.
% }

\section{Evaluation}
\label{sec:eval}

Our evaluation considers the following three research questions:
\begin{itemize}
\item [RQ1] {\bf Effectiveness} of \name{}: How effectively can
  \name\ synthesize complex refinement type queries? How does
  \name\ compare against other specification-guided, component-based
  synthesis tools in its ability to synthesize programs when queries
  define fine-grained constraints?

% How do other tools behave on these queries, here basically create a plot with x axis
% being the time (nromalized to one for each query) and for each tools we vary the number of
% i/o examples: 1 3 6 and see how close they come to Hegel.
% \item [RQ2] {\bf Comparison} to other approaches: How does \name\
%   compare against other specification-guided, component-based
%   synthesis tools in its ability to synthesize programs when queries
%   express fine-grained constraints?

\item [RQ2] {\bf Scaling} to query complexity: Complex queries
  correlate to the size and control-flow complexity in synthesized
  outputs. How effective is \name\ in synthesizing programs as query
  complexity scales?
  
  %% How effectfive \name is to synthesize programs of medium sizes (8-20 call sequences) and complex control flows which are beyond the capabilities of the state-of-the-art type/specification-guided CBS syntheses? 
    % \AM{A new set of becnhmarks, beyond any synthesizer, again, we also present the number of terms enumerated}

\item [RQ3] {\bf Impact} of QTA reduction strategies on synthesis
  efficiency and quality: How significant are the benefits of pruning
  and similarity reductions in reducing QTA size and the search space
  over which the synthesis procedure operates?
  
\end{itemize}

\subsection{Benchmarks}

To effectively answer RQ[1-3], we required component-based synthesis
benchmarks of the kind shown in Section~\ref{motivation}. To this end,
we have collected a set of Hoogle+~\cite{hoogleplus, digging-fold} and
Hectare~\cite{ecta} benchmarks originally used for (simple)
type-guided synthesis over Haskell libraries, and re-implemented them
in OCaml for use by \name.  \ADD{To avoid selection bias towards
  queries with only a particular kind of features (e.g., Hoogle+
  queries are primarily first-order while Hectare mostly avoids
  first-order queries), we divided the original benchmarks from these
  two sources into three main categories: \textit{standard}
  first-order queries primarily from Hoogle+; \textit{higher-order}
  queries primarily from Hectare and \textit{polymorphic} queries from
  both these sources.}
  
% \AM{In previous submission, the reviewers saw the number of unrefined queries and got confused that we are performing better, we should find a way to address this.}
% We compare the performance of \name{} with three tools, Hoogle+~\cite{hoogleplus, digging-fold} (a type-guided cbs using i/o examples), Hectare~\cite{ecta} (A type-guided synthesis using tree automata) and Synquid~\cite{synquid} (A Refinement type-guided synthesis with nai\`ve enumeration).

%To facilitate a comparison between \name{} and these tools, 

To create realistic, refined queries from these, and compare \name{}
against other tools, we refine each of these to include three
different user-provided refinements.  These refinements are carefully chosen so
that they \emph{invalidate} the synthesized output on the original
(unrefined) query.
%% Hoogle+ allows input-output
%% examples to capture such refinements, while \name{} and
%% Synquid~\cite{synquid} requires users to specify such refinements as
%% type-based specifications.  
We specify the intent of these refinements in the Hoogle+ runs using 4
I/O examples; Hectare does not accept query refinements or I/O
examples and performs synthesis only based on the (non-refined)
standard types given in the library.  To make the comparison as fair
as possible, we drop queries in our benchmark set that cannot either
be easily refined using I/O examples, or where the refinement type
specification was trivial compared to the I/O example.  Our goal was
to choose queries that collectively cover all three categories with
non-trivial refined queries, and with a similar degree of complexity
in both I/O and refinement type specifications.  These requirements
led us to select 14 queries across the three categories that we found
to be amenable for such comparison.~\footnote{
 See supplemental material for a detailed description of our benchmark
  set, along with one example with explanation.} We then compare the performance of Hoogle+,
Synquid~\cite{synquid}, a deductive synthesis tool that uses
refinement types for its specification, and \name{} on 42 (14 $\times$
3) different refined queries, making our comparison extensive.
Table~\ref{tab:results-rq1}, shows these original queries type
specifications along with description of each of the refined queries.

\subsection{RQ1: Effectiveness and Comparison with Other tools}
\paragraph{Results}
%Define the table
\NEW{The table in Figure}~\ref{tab:results-rq1} presents the main experimental results for
RQ1. 
Tables in Fig 4 and 5 in [1] show a) the number of SMT calls, b) time spent for constraints solving, c) Number of original QTA states (|Q|), and d) QTA states after minimization (|Q| min) for RQ1 and RQ2  benchmarks respectively. 
\NEW{The first column gives a symbolic name for the refined
benchmark derived from the original Hoogle+ and ECTA benchmarks.} For
example, the \S{Nth1} benchmark requires a solution that performs a
non-trivial swapping of the values at given indexes in the input
list~\footnote{See supplemental material for details on the original simply-typed queries and a description of the refinements for each benchmark}. 
Several benchmarks also include higher-order functions; these
are given as the last five ($\times$ three benchmarks) set of queries,
marked with \textdagger.

The input to \name{} is a refinement-type query and a pre-annotated
set of libraries. To quantify query specification burden, the next
column in the table provides a measure of specification size in terms
of the number of conjunctions($\wedge$) and disjunctions($\vee$). The
average size of these specs is around four-five conjuncts making this
overhead small, particularly when compared against size of the
library, the complexity of the search, and the number of examples
needed to capture these in i/o settings like Hegel.  The next three
columns show the results for running \name{}(\S{He}), Hoogle+(\S{H+})
and Synquid (\S{Sn}) on each benchmark. For each of these runs, we
used a timeout limit of three minutes and maximum term size bound of
five library function calls.  All results were performed on a standard
off-the-shelf laptop with 8GB of memory.

As the table shows, \name{} is able to solve all benchmarks under 11
seconds, with an average time of around 7.6 seconds.  Hoogle+ (H+),
was able to solve only a fraction (6/42) of the benchmarks, 
taking approximately \textbf{6x} more time to yield a solution than
\name.

%% Varying the number of
%% examples (from 2 to 6) yielded no change in results. This is
%% understandable as it performs a simple type-guided synthesis, with
%% post filtering using i/o examples, thus, increasing the number of
%% examples does not directly effect the search process.
There is no data to report for Hectare because it does not support
logical refinements (either as types or I/O examples) on queries.
Column `Sn` shows the synthesis time for Synquid, which solves around
20/42 of these benchmarks, while taking substantially more time,
around 4.5x, on average.  This is understandable, given that Synquid's
goal is not efficient CBS and hence does not employ any efficient
pruning or search mechanism.

\NEW{The next four columns provide statistics for the QTA reduction and constraint solving overhead in \name{}. These provide a) the number of SMT calls, b) time spent for constraint solving, c) the number of original QTA states (|Q|), and d) the number of QTA states remaining after minimization (|Q| min) for RQ1 and RQ2  benchmarks, resp. Overall, on average, Hegel makes approximately 30 SMT calls per benchmark while spending close to 3.8 seconds on average per benchmark; thus, approximately 40\% of overall synthesis time is spent in constraint solving.}  The final column (\#S) gives the size of \name's solution in terms of the
number of library function calls in the synthesized result. These
numbers, in most cases, differ from the solutions for the original
query. In fact, in several cases the solutions also introduce new
control-flow branches (2 branches in each case) that were absent in
the original; we label these benchmarks with a $\star$.

{\scriptsize
\begin{figure}[h]
\begin{tabular}{| l | l | l | l | l | l | l | l | l | l | }
\hline
\textbf{Name} & \#$\wedge$/$\vee$ & \multicolumn{3}{|c|}{\textbf{Refined Time(s)/\#}} & \multicolumn{4}{|c|}{\textbf{QTA \& SMT Stats}} & \#S \\
\hline
 &  & He & H+ & Sn & \#SMT & SMT Time & $|Q|$ & $|Q|_{\min}$ &  \\

\hline
Nth1   & 4 & 5.1 & 39.4 & 21.3 & 22 & 3.07 & 239 & 53 & 4\\
Nth2   & 3 & 7.6 &       & 22.7 & 25 & 3.40 & 199 & 56 & 5 \\
Nth3   & 4 & 9.1 &       &      & 33 & 4.01 & 210 & 78 & 5 \\
\hline
RevApp1	& 3 & 5.1 & 49.5 &      & 18 & 2.87 & 110 & 56 & 3 \\
RevApp2	& 3 & 7.3 &       &      & 22 & 3.90 & 166 & 61 & 4 \\
RevApp3  & 4 & 5.4 &       &      & 18 & 3.03 & 201 & 43 & 4 \\
\hline
RevZip1  & 3 & 6.1 & 43.2 & 22.6 & 31 & 4.21 & 278 & 69 & 4 \\
RevZip2  & 4 & 6.9 &       & 28.1 & 21 & 3.80 & 189 & 69&  5 \\
RevZip3  & 4 & 8.5 &       &      & 39 & 5.07 & 301 & 62 & 6\\
\hline
SplitAt1 & 4 & 6.8 &       & 32.1 & 19 & 2.10 & 156 & 47 & 5 \\
SplitAt2 & 5 & 7.3 &       & 24.3 & 28 & 4.13 & 176 & 76 & 4 \\
SplitAt3 & 4 & 7.8 &       &      & 25 & 4.40 & 173 & 54 & 5 \\
\hline
Nth\_Incr1 & 3 & 5.2 & 35.4 & 36.8 & 19 & 3.70 & 143 & 38 & 3 \\
Nth\_Incr2 & 4 & 7.4 & 56.3 &      & 37 & 5.12 & 187 & 87 & 5\\
Nth\_Incr3 & 4 & 7.8 &       & 39.5 & 26 & 4.30 & 165 & 73 & 5\\
\hline
CEdge1 & 4 & 6.4 &       & 21.5 & 21 & 3.10 & 110 & 39 & 4 \\
CEdge2 & 4 & 5.2 &       & 23.7 & 11 & 2.87 & 145 & 34 & 3 \\
CEdge3 & 5 & 8.6 &       &      & 24 & 2.98 & 167 & 42 & 5 \\
\hline
AppendN1 & 4 & 6.0 & 47.7 &      & 21 & 3.04 & 267 & 61 & 4\\
AppendN2 & 4 & 10.4 &     &      & 39 & 4.70 & 310 & 76 & $7^{\star}$ \\
AppendN3 & 4 & 7.5 &      & 25.8 & 31 & 4.60 & 234 & 57 & 5 \\
\hline
SplitStr1 & 4 & 6.5 &     &      & 17 & 3.10 & 212 & 41 & 5\\
SplitStr2 & 4 & 5.2 &     & 36.2 & 18 & 1.90 & 245 & 63 & 5\\
SplitStr3 & 6 & 7.1 &     &      & 21 & 3.00 & 277 & 87 & 5\\
\hline
LookRange1 & 4 & 8.2 &     & 43.6 & 37 & 2.80 & 413 & 99 & $6^{\star}$ \\
LookRange2 & 6 & 9.1 &     &      & 65 & 3.30 & 512 & 114 & $7^{\star}$\\
LookRange3 & 4 & 8.6 &     &      & 57 & 3.80 & 511 & 114 & $7^{\star}$\\
\hline
$\S{Map1}^{\textnormal{\textdagger}}$  & 5 & 7.5 &     &      & 41 & 3.60 & 219 & 110 & 6 \\
$\S{Map2}^{\textnormal{\textdagger}}$  & 6 & 5.2 & 45.1 & 34.1 & 25 & 2.20 & 198 & 55 & 4 \\
$\S{Map3}^{\textnormal{\textdagger}}$ & 5 & 5.0 & 33.5 & 29.8 & 22 & 1.40 & 176 & 53 & 4 \\
\hline
$\S{MapDouble1}^{\textnormal{\textdagger}}$ & 4 & 10.8 &     &      & 36 & 3.02 & 399 & 89 & 5 \\
$\S{MapDouble2}^{\textnormal{\textdagger}}$ & 4 & 8.9 &      &      & 51 & 3.60 & 465 & 88 & 4  \\
$\S{MapDouble3}^{\textnormal{\textdagger}}$ & 6 & 8.4 &      & 42.8 & 55 & 2.60 & 452 & 101 & 4\\
\hline
$\S{ApplyNAdd1}^{\textnormal{\textdagger}}$ & 6 & 7.9 &      &      & 27 & 3.10 & 259 & 76 & 5\\
$\S{ApplyNAdd2}^{\textnormal{\textdagger}}$ & 6 & 6.9 & 63.1 & 34.2 & 17 & 3.60 & 331 & 59 & 5  \\
$\S{ApplyNAdd3}^{\textnormal{\textdagger}}$ & 6 & 9.1 &      &      & 34 & 5.20 & 323 & 79 & 6  \\
\hline
$\S{ApplyNInv1}^{\textnormal{\textdagger}}$ & 5 & 9.5 &      &      & 34 & 4.80 & 299 & 83 & 5 \\
$\S{ApplyNInv2}^{\textnormal{\textdagger}}$ & 5 & 7.6 &      & 32.4 & 11 & 6.30 & 134 & 38 & 5\\
$\S{ApplyNInv3}^{\textnormal{\textdagger}}$ & 6 & 8.1 &      & 55.6 & 13 & 4.80 & 156 & 41 & 5 \\
\hline
$\S{ApplyList1}^{\textnormal{\textdagger}}$ & 4 & 12.3 &     &      & 65 & 7.50 & 519 & 132 & $8^{\star}$\\
$\S{ApplyList2}^{\textnormal{\textdagger}}$ & 4 & 10.5 &     &      & 59 & 7.70 & 483 & 122 & $8^{\star}$\\
$\S{ApplyList3}^{\textnormal{\textdagger}}$ & 5 & 7.9 &      & 42.6 & 43 & 4.10 & 277 & 75 & $6^{\star}$\\
\hline
\end{tabular}
\caption{Results for experiments with  Refined Hoogle+ (H+) and ECTA benchmarks.  The details of the original queries and a description of the refinements for each benchmark is given in the supplemental material.}
\label{tab:results-rq1}
\end{figure}
}

% {\scriptsize
% \begin{figure}[h]
% %name | refined time Hegel| | refined terms Hegel | \#C | \#B |\#C-M | \#B-M
% \begin{tabular}{| l | l | >{\columncolor[gray]{0.8}}l | l | l | l | l | l | l | l | l |}
% \hline
% % \multicolumn{3}{|c|}{BS1, Sparsified Hoogle+ Benchmarks}  \\ 
% %  \hline
% Name & Desc.  & \#$\wedge$/$\vee$ & T(He)  & \#C & \#B  & \#R & \# {T(Sn)} \\
% \hline

% NLInsert & Add a newsletter and user  &  6 &   22.7	                & 16  & 2   &    31
%  & 	126.2\\
% NLRemove & Remove a newsletter and user & 4 &    39.9	            & 20 & 4	  &35
%  &   \_ \\
% NLR\_Remove	 & Read articles list and remove & 5  &  42.8	 	    &19  & 4   & 21
%  & \_ \\
% NLInv	&  Remove with uniqueness invariant & 8  &  52.5	   	       & 25	& 4	  & 36
% & \_ \\
% FWInsert&	Insert a normal device  & 4 &  31.2	& 15	 & 2	 & 33
%  &	124.6 \\
% FWMkCentral& Insert a central device & 4	&   65.2		       & 33  &4	 & 53  &   \\   
% FWInsConn& 	Insert a device connected to all & 5  &  36.8      & 14
%     & 2	 &  59	&   \\   
% FWInvert& Invert the connections & 4	 &33.9	      & 14	    &	4  & 47  &   \\   
% FWInvertDel& Delete, and invert connections & 6 & 47.3		    & 	17    & 4  & 38  &   \\   

% \hline

% \end{tabular}
%  \caption{Results for tailored specification-guided synthesis benchmarks, The \#C and \#B gives the total number of function calls and branches in the synthesized solution. \#R gives the number of transitions \name{} merged during the Similarity reduction and Irrelevant code pruning phases.}
%  \label{tab:results-rq2}
 
%  \end{figure}
%  }

{\scriptsize
\begin{figure}[h]
\begin{tabular}{| l | p{3cm} | >{\columncolor[gray]{0.8}}l | l | l | l | l | r | r | r | r | r |}
\hline
\textbf{Name} & Desc. & \#$\wedge$/$\vee$ &  \multicolumn{4}{|c|}{\textbf{Results Hegel}} & \multicolumn{4}{|c|}{\textbf{QTA \& SMT Stats}} & T(Sn) \\
\hline
 &  &  & T(He)  & \#C & \#B  & \#R & \#SMT & SMT(s) & $|Q|$ & $|Q|_{\min}$ &  \\
\hline
NLInsert     & Add a newsletter and user           & 6 & 22.7 & 16 & 2 & 31 & 110 & 11.12 & 779  & 212 & 126.2 \\
NLRemove     & Remove a newsletter and user        & 4 & 39.9 & 20 & 4 & 35 & 189 & 19.34 & 1201 & 372 & \_     \\
NLR\_Remove  & Read articles list and remove       & 5 & 42.8 & 19 & 4 & 21 & 213 & 18.20 & 1331 & 381 & \_     \\
NLInv        & Remove with uniqueness invariant     & 8 & 52.5 & 25 & 4 & 36 & 154 & 24.19 & 1398 & 435 & \_     \\
FWInsert     & Insert a normal device               & 4 & 31.2 & 15 & 2 & 33 & 166 & 12.34 & 945  & 298 & 124.6 \\
FWMkCentral  & Insert a central device              & 4 & 65.2 & 33 & 4 & 53 & 259 & 27.13 & 1611 & 401 &        \\
FWInsConn    & Insert a device connected to all     & 5 & 36.8 & 14 & 2 & 59 & 218 & 15.12 & 806  & 261 &        \\
FWInvert     & Invert the connections               & 4 & 33.9 & 14 & 4 & 47 & 197 & 15.70 & 1176 & 323 &        \\
FWInvertDel  & Delete, and invert connections       & 6 & 47.3 & 17 & 4 & 38 & 184 & 22.90 & 1352 & 421 &        \\
\hline
\end{tabular}

\caption{Results for tailored specification-guided synthesis benchmarks, The \#C and \#B gives the total number of function calls and branches in the synthesized solution. \#R gives the number of transitions \name{} merged during the Similarity reduction and Irrelevant code pruning phases.}
\label{tab:results-rq2}
 
\end{figure}
}
 
% \AM{Do not highlight cobalt too much rather talk about refinement typed queries from sources like Hoogle, Cobalt, rather than using the query, use the output generated by the solution.}

\subsection{RQ2, Scaling \name{} to larger and complex queries}
To answer RQ2, we consider synthesis query benchmarks whose solutions
require longer call sequences and more complex control flows than the
queries given in the table shown in Figure~\ref{tab:results-rq1}.  We
have collected and refined eight queries adapted from verification
benchmarks~\cite{database-examples} that are particularly amenable for
CBS techniques.  Figure~\ref{tab:results-rq2} lists these benchmarks,
which are defined over two database
applications~\cite{database-examples}. We only show \name{} and
Synquid results because Hoogle+ was not able to produce solutions
within the timeout bound (3 minutes).

The first is a {\it Newsletter} database (queries with names NL\ldots
in the table) that defines a single table \S{NS} equipped with various
attributes (e.g., \emph{newsletter, user, subscribed, articles, code,
  clear\_email, add\_email}, etc.).  For example, query 
    \S{NLRRemove} is shown as a comment in Figure~\ref{fig:synthesized}, it encodes the following problem: \textit{Synthesize a
    function that returns a list of articles for a given newsletter
    ({\sf n}) and a given user ({\sf u}) in a database ({\sf d}),
    along with an updated database that does contain {\sf u} and {\sf
      nl}, while keeping the email address if {\sf u} has opted for promotions}.  The
  second application implements a network firewall database (queries
  with names FW\ldots in the table) that manages two tables, a table
  of devices and a table storing sender-receiver links.  These
  benchmarks are adapted from their original definition to employ a
  monadic style that threads database state through calls.
  
  Synthesizing programs from queries of this kind must take into
  account appropriate protocols associated with the libraries, e.g.,
  to establish a connection to a device that is not currently in the
  device table requires that the device first be added. These
  constraints also require conditional-control flows in the solutions.

\paragraph{Results}
The first two columns in Figure~\ref{tab:results-rq2} give the name
and a small high-level description of the benchmark.  The next two
columns `He' gives the synthesis time in seconds. \name{} succeeds in
finding solutions to all queries with a synthesis time in the range of
22.7 seconds to little more than a minute.  Its ability to solve these
queries can be attributed to an efficient exploration enumeration
strategy that effectively reduced, either via {\sc Prune} or {\sc
  Similarity} reduction strategies, a large number of tranistions
during synthesis. These numbers are given in the (\#R) column and
ranges from 21 in {\sc NLR\_Remove}, to as high as 59 transitions in
{\sf FWInsConn}.  Note that the effective actual saving in enumeration
is much more that this, as an exponential number of other transitions
built over these are never constructed.

To quantify the complexity of the synthesized solutions, we also list
the properties of the solutions synthesized in terms of the total
number of function calls (i.e., the size of the solution) as well as
solution complexity in terms of the number of control flow branches.
For instance, given the query {\sf NLRRemove}, the challenge is to
synthesize a solution that maintains a specific contract associated
with each library function; these include the requirement that a) the
user must be unsubscribed before removal, b) if the user has not opted
for \textit{promotions}, the email for the user must be cleared, etc.
Figure~\ref{fig:synthesized} shows the synthesized program generated
by \name{} for this query.  Note that the solution includes a total of
19 function calls and exhibits complex control flows (4 branches).
\NEW{The next four columns show: a) the number of SMT calls, b) the time spent for constraint solving, c) the number of original QTA states (|Q|), and d) the number of QTA states remaining after minimization (|Q| min). The average number of QTA states produced is around 1200, while Hegel's pruning and minimization results in a roughly 70\% reduction.  The last column shows synthesis times for Synquid, which is the only
other tool that can work with refined specifications. It produces results for only the smallest of the
benchmarks, taking a little over two minutes on those, while timing-out
on all other cases.}\\
\begin{wrapfigure}{l}{.48\textwidth}
\vspace*{-.25em}
\begin{minted}[fontsize = \scriptsize, escapeinside=&&,linenos]{ocaml}
(*nLRRemove : (n : nl) -> (u : user) -> 
  (d : {v: [nlrecord] | mem (v , n , u)}) -> 
  {v : (  f : article * s : [nlrecord]) |  
  mem (f, articles (s)) 
  &$\wedge$&  &$\neg$& nlmem (s, n, u) 
  &$\wedge$& (promotions (s, u) => email (s, u))
  }*)
fun n u d ->
    let x = read (d, n, u) in 
    let x0 = fst (x) in 
    let d0 = snd (x) in
    let d1 = confirmU (d0, n, u) in
    let x1 = promotions (d1, n, u) in 
    if (not (x1)) then 
        let subscribed = subscribed (d1, n, u) in 
        if (length (subscribed) > 0) then 
            let d2 = clear_email (d1, n, u) in 
            let d3 = unsubscribe (d2, n, u) in 
            let d4 = remove (d3, n, u) in 
            (x0, d4)
        else 
            let d5 = unsubscribe (d1, n, u) in 
            let d6 = remove (d5, n, u) in 
            (x0, d6)
            
    else 
        let d7 = unsubscribe (d1, n, u) in 
        let d8 = remove (d7, n, u) in 
        (x0, d8)
\end{minted}
\caption{Synthesized Program for NLR\_Remove}
\label{fig:synthesized}
\end{wrapfigure}

    \subsection{RQ3: Impact of irrelevant code pruning and similarity reduction}
Because RQ3 cuts across both set of benchmarks, we perform several
ablation experiments over the queries described in the
Figures~\ref{tab:results-rq1} and ~\ref{tab:results-rq2}.  We create
three variants of \name{}, \textit{viz.} (i) \textsf{Hegel(-P)}, a
QTA-based synthesis implementation without the irrelevant code
reduction (i.e. comment out the {\sc Prune} call at line 7 in
Algorithm~\ref{alg:enumerate}), but retaining \textit{similarity
  reduction}; (ii) \textsf{Hegel(-S)}, a variant of \name{} with
support for pruning but \textit{without} similarity reduction (i.e.,
lines 8 and 9 in Algorithm~\ref{alg:enumerate}) are commented out);
and, (iii) {\sf Hegel(-All)}, a baseline variant that constructs the
QTA without performing any reduction (i.e., removes lines 7-9 in the
algorithm).  We compare these variants in terms of two main metrics,
overall \textit{synthesis times} and the size of the search space in
each case after the reduction, shown by \textit{number of program
  terms enumerated} during search, compared to the base-line ({\sf
  Hegel(-All)}.

\begin{figure}[t!]

\centering 
\includegraphics[width=1.0\textwidth]{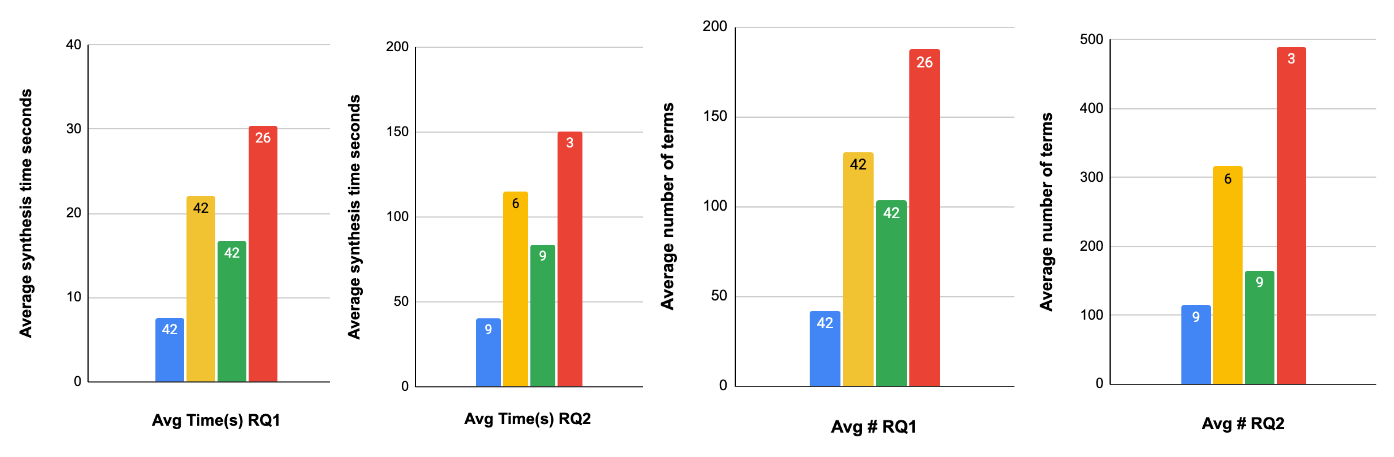}
\caption{Comparison of Hegel and its variants {\color{blue}{\sf
      Hegel}}, {\color{amber}{\sf Hegel(-S)}}, {\color{bargreen}{\sf
      Hegel(-P)}}, {\color{red}{\sf Hegel(-ALL)}}, on average synthesis times
  and the number of candidate terms generated on RQ1 and RQ2. The
  labels on each bar show the number of benchmarks solved by these
  variants out of 42 in the case of RQ1 and 9 in the case of RQ2.}
\label{fig:refined-time-rq1}
\end{figure}
%\paragraph{Results}

The first two charts in Figure~\ref{fig:refined-time-rq1} show results for overall average
synthesis times across the two sets of benchmark queries described
earlier.  We note that both {\sf Hegel(-S)}, and {\sf Hegel(-P)} can solve
all queries from RQ1, but at a cost which is 2 - 3X greater than
\name{}. {\sf Hegel(-All)} on the other hand fails on almost half of the benchmarks.  
In contrast,
although {\sf Hegel(-P)} was also able to solve the full complement of
queries studied in RQ2, it did so with a considerable larger overhead
compared to \name{}, while here the the irrelevant code pruning ({\sf Hegel(-S)}) alone is insufficient to scale the variant to these challenging benchmarks and it fails to solve 3/9 benchmarks. 
The second pair of charts and show the average number of terms
enumerated by these different variants, showing the reduction of search space by each reduction strategy, with {\sf Hegel(-All)} as the baseline.  Here we see, with the combined reduction strategies, {\sf Hegel} sees the maximum search space reduction, while the other two variants {\sf Hegel(-P)} and {\sf Hegel(-S)} having much larger search spaces, (anywhere from 2-4.5X more) without necessarily solving the same number of queries.

\section{Related Work}
\label{sec:related}

\noindent 
\emph{Component-based Synthesis.}  There is a long line of work on the
use of CBS in the context of domain-specific
languages~\cite{table-synthesis, oracle-guided-synthesis} as well as
general-purpose programming
domains~\cite{sypet,tygus,cdcl-synthesis,rbsyn,frangel,viser,cobalt-tech}.
Despite various technical differences, these approaches all include
some form of search at their core, which they tackle using basic type
information~\cite{tygus, sypet}, or a combination of types and limited
effects~\cite{rbsyn}.  Our contributions in this paper extend prior
work by enabling CBS to be applied when specifications and queries are
equipped with logical refinements, substantially increasing the
complexity of the search process.  While some prior
work~\cite{cobalt-tech, cdcl-synthesis} also use logical properties to
prune the feasible search space, they provide only limited benefits in
the absence of a data structure like QTA with respect to query
complexity scaling.  For example, running~\cite{cobalt-tech} on the
benchmarks in Figure~\ref{tab:results-rq2} produces results similar to
Hegel(-ALL).

\emph{Using similarity/equality information for search.}  Recent
interest in utilizing equality information, and using ways to
calculate equality saturation sets using e-graphs~\cite{Egg} allows
efficient reduction of an enumeration space similar to the motivation
underlying our approach. Equality saturation has been applied to
enable efficient abstraction learning~\cite{babble}
inductive synthesis~\cite{egg-synthesis} and program
analysis~\cite{egg-analysis}. However, all these works depend
crucially on fast calculation of the equality saturation set, making
them suitable for syntactic similarity relations, but it is not
obvious how to extend this technique to provide support for semantic
(subtype-based) similarity checks as in our setting.  Nonetheless, we
leave as future work, questions related to how QTA-like structures can
benefit from the general capabilities provided E-graphs.  Another line
of work in this vein~\cite{manual} asks users to provide logical
equivalences between operators that can be then used to accelerate the
synthesis process. This requirement has utility when dealing with
small DSLs where a user is likely to understand operator semantics and
their equivalences, but we expect will be infeasible in a
component-based synthesis setting that deals with libraries with a large
number of methods.

Our idea of similarity reduction is also related to the notion of
\textit{observational equivalence} found in programming-by-example
synthesis
approaches~\cite{Albarghouthi2013,Miltner2022,Dillig23};
  these techniques compare synthesized programs on a given set of
  inputs and prune the search-space in a bottom-up, inductive
  synthesis setting.  Its inherent unsoundness makes this mechanism
  infeasible for specification-guided synthesis.

\emph{Tree automata for program synthesis.} As described earlier, prior
work has leveraged tree automata to compactly capture the space of
programs~\cite{dace,ecta,reactive} in a synthesis setting; such
automata have also been used in the synthesis of reactive
systems~\cite{reactive}. In fact, recent work~\cite{ecta} has also
proposed techniques to qualify the states of such automata, albeit
with syntactic equality constraints~\cite{tata,Reduction} to capture
dependences between sub-spaces, allowing efficient reduction of the
search space over base types.  We see our contribution as a
continuation of these efforts, generalizing them to handle richer
semantic dependences expressed in specifications and queries.

Closely related to our work is Blaze~\cite{blaze} which introduces an
abstract finite tree automata (AFTA) to help implement its synthesis
procedure.  AFTA's states, like QTA's, are also defined by a type plus
a predicate. However, there are a number of important differences
between the two approaches; first, unlike the qualifiers in QTA, which
are refinement types, the abstract predicates in AFTA are formulas
capturing an abstract domain~\cite{neilson}.  The annotations on the
states are much more complex in QTAs, expressed as quantified formulas
over program variables; in contrast, AFTA's predicates are
quantifier-free formulas over an abstract domain.  QTA qualifiers, on
the other hand, are used to navigate a compact representation of the
search space and to help find similarities between subspaces in a
refinement-type guided deductive synthesis procedure.  Second, unlike
AFTA's, QTAs also support function types, higher-order programs, and
let bindings.  These differences lead to substantially different
synthesis algorithms in these two approaches.  CTA~\cite{cata} also
adds constrains to traditional tree automata similar to how symbolic
finite automata extends finite automata. Particularly, CTA allows
transitions guarded by a logical formulas from a decidable theory,
allowing them to effectively capture relational properties with
deciadable acceptance checking. Unfortunately, these logical formulas
cannot relate sub-automata, like in QTA or ECTA, thus, making them
less effective to capture typing semantics. Allowing CTA-like
constraints in QTA, however, may allow us to extend our synthesis
approach to yield recursive and mutual recursive functions; we leave
this as part of future work..

\emph{Refinement types and conflict-driven learning for synthesis.}
Several earlier works have used refinement types for program
synthesis~\cite{cobalt-tech, synquid, ecta}. Although, Synquid's
synthesis is also guided by Refinement Types, its goals differ from
ours, and it does not aim towards efficient CBS in refinement-typed
space. A family of works using CDCL may also be somewhat related,
however, the CDCL avoidance~\cite{cobalt-tech, sypet} of equivalent
failing terms alone has limited pruning capabilities, as the
synthesizer still has to explore a large number of terms that have
neither failed (they may still lead to a correct solution when further
explored) nor reached the maximum term-size depth. In such cases, a
CDCL-based synthesizer is still forced to explore many equivalent
(non-failed) terms; e.g, all the terms shown in blue, gray or yellow
boxes in Figure~\ref{fig:explosion} may still be generated using such
techniques. Roughly speaking, the overall performance of such an
approach is on par with Hegel(-S), the green bar in
Figure~\ref{fig:refined-time-rq1}.

%% \TODO{See if there are any other verticals for comparison}
%% \paragraph{Specifications for synthesis} 
%% \TODO{\ldots}

% \TODO{Synthesizing Reactive Programs: P. Madhusudan
% uses tree automata for for synthesizing reactive programs.}

% \TODO{Look into each and do a detailed comparison as well, but first the earlier figure.}

\section{Conclusions}
\label{sec:conc}

This paper describes a new component-based synthesis algorithm and
tool (\name) designed to operate on libraries and queries that are
equipped with refinement-type specifications.  These specifications
can impose significant constraints on the set of feasible solutions
making na\"{\i}ve enumeration of the search space impractical.  We
propose a new tree automata variant (QTA) to succinctly represent the
search space in this setting, and propose a number of semantics-based
optimizations to greatly reduce search overhead.  allow efficient
construction and enable semantic-based similarity checking among
candidate terms to greatly reduce search overhead.  Our experimental
results demonstrate that \name{} is able to successfully synthesize
correct outputs given complex query inputs over a range of application
benchmarks that exceed the capabilities of existing systems.

\bibliographystyle{plain}

\section*{Data Availability Statement}

Our supplementary material includes an anonymized artifact. This artifact
contains the OCaml source code for \name{} and our
suite of benchmark programs. We intend to submit this artifact with additional scripts to automatically generate the results for evaluation by
the artifact evaluation committee should this paper be accepted.

\bibliography{paper}


\begin{thebibliography}{36}

%%% the Web address.

\ifx \showCODEN    \undefined \def \showCODEN     #1{\unskip}     \fi
\ifx \showDOI      \undefined \def \showDOI       #1{#1}\fi
\ifx \showISBNx    \undefined \def \showISBNx     #1{\unskip}     \fi
\ifx \showISBNxiii \undefined \def \showISBNxiii  #1{\unskip}     \fi
\ifx \showISSN     \undefined \def \showISSN      #1{\unskip}     \fi
\ifx \showLCCN     \undefined \def \showLCCN      #1{\unskip}     \fi
\ifx \shownote     \undefined \def \shownote      #1{#1}          \fi
\ifx \showarticletitle \undefined \def \showarticletitle #1{#1}   \fi
\ifx \showURL      \undefined \def \showURL       {\relax}        \fi
% invisible to TeX
\providecommand\bibfield[2]{#2}
\providecommand\bibinfo[2]{#2}
\providecommand\natexlab[1]{#1}
\providecommand\showeprint[2][]{arXiv:#2}

\bibitem[Albarghouthi et~al\mbox{.}(2013)]%
        {Albarghouthi2013}
\bibfield{author}{\bibinfo{person}{Aws Albarghouthi}, \bibinfo{person}{Sumit
  Gulwani}, {and} \bibinfo{person}{Zachary Kincaid}.}
  \bibinfo{year}{2013}\natexlab{}.
\newblock \showarticletitle{Recursive Program Synthesis}. In
  \bibinfo{booktitle}{\emph{Computer Aided Verification}},
  \bibfield{editor}{\bibinfo{person}{Natasha Sharygina} {and}
  \bibinfo{person}{Helmut Veith}} (Eds.). \bibinfo{publisher}{Springer Berlin
  Heidelberg}, \bibinfo{address}{Berlin, Heidelberg},
  \bibinfo{pages}{934--950}.
\newblock
\showISBNx{978-3-642-39799-8}


\bibitem[Biere et~al\mbox{.}(2009)]%
        {cdcl-sat}
\bibfield{editor}{\bibinfo{person}{Armin Biere}, \bibinfo{person}{Marijn
  Heule}, \bibinfo{person}{Hans van Maaren}, {and} \bibinfo{person}{Toby
  Walsh}} (Eds.). \bibinfo{year}{2009}\natexlab{}.
\newblock \bibinfo{booktitle}{\emph{Handbook of Satisfiability}}.
  \bibinfo{series}{Frontiers in Artificial Intelligence and Applications},
  Vol.~\bibinfo{volume}{185}. \bibinfo{publisher}{{IOS} Press}.
\newblock
\showISBNx{978-1-58603-929-5}


\bibitem[Bowers et~al\mbox{.}(2023)]%
        {egg-synthesis}
\bibfield{author}{\bibinfo{person}{Matthew Bowers}, \bibinfo{person}{Theo~X.
  Olausson}, \bibinfo{person}{Lionel Wong}, \bibinfo{person}{Gabriel Grand},
  \bibinfo{person}{Joshua~B. Tenenbaum}, \bibinfo{person}{Kevin Ellis}, {and}
  \bibinfo{person}{Armando Solar-Lezama}.} \bibinfo{year}{2023}\natexlab{}.
\newblock \showarticletitle{Top-Down Synthesis for Library Learning}.
\newblock \bibinfo{journal}{\emph{Proc. ACM Program. Lang.}}
  \bibinfo{volume}{7}, \bibinfo{number}{POPL}, Article \bibinfo{articleno}{41}
  (\bibinfo{date}{jan} \bibinfo{year}{2023}), \bibinfo{numpages}{32}~pages.
\newblock
\urldef\tempurl%
\url{https://doi.org/10.1145/3571234}
\showDOI{\tempurl}


\bibitem[Cao et~al\mbox{.}(2023)]%
        {babble}
\bibfield{author}{\bibinfo{person}{David Cao}, \bibinfo{person}{Rose Kunkel},
  \bibinfo{person}{Chandrakana Nandi}, \bibinfo{person}{Max Willsey},
  \bibinfo{person}{Zachary Tatlock}, {and} \bibinfo{person}{Nadia
  Polikarpova}.} \bibinfo{year}{2023}\natexlab{}.
\newblock \showarticletitle{Babble: Learning Better Abstractions with E-Graphs
  and Anti-Unification}.
\newblock \bibinfo{journal}{\emph{Proc. ACM Program. Lang.}}
  \bibinfo{volume}{7}, \bibinfo{number}{POPL}, Article \bibinfo{articleno}{14}
  (\bibinfo{date}{jan} \bibinfo{year}{2023}), \bibinfo{numpages}{29}~pages.
\newblock
\urldef\tempurl%
\url{https://doi.org/10.1145/3571207}
\showDOI{\tempurl}


\bibitem[Chargu\'eraud et~al\mbox{.}(2017)]%
        {vocal}
\bibfield{author}{\bibinfo{person}{Arthur Chargu\'eraud},
  \bibinfo{person}{Jean-Christophe Filli{\^a}tre}, \bibinfo{person}{M{\'a}rio
  Pereira}, {and} \bibinfo{person}{Fran\c{c}ois Pottier}.}
  \bibinfo{year}{2017}\natexlab{}.
\newblock \bibinfo{title}{{VOCAL} -- {A} {V}erified {OC}aml {L}ibrary}.
\newblock \bibinfo{howpublished}{ML Family Workshop}.
\newblock


\bibitem[Comon(1997)]%
        {tata}
\bibfield{author}{\bibinfo{person}{Hubert Comon}.}
  \bibinfo{year}{1997}\natexlab{}.
\newblock \showarticletitle{Tree automata techniques and applications}.
\newblock
\urldef\tempurl%
\url{https://api.semanticscholar.org/CorpusID:2092186}
\showURL{%
\tempurl}


\bibitem[Dauchet et~al\mbox{.}(1995)]%
        {Reduction}
\bibfield{author}{\bibinfo{person}{Max Dauchet}, \bibinfo{person}{Anne-Cécile
  Caron}, {and} \bibinfo{person}{Jean-Luc Coquidé}.}
  \bibinfo{year}{1995}\natexlab{}.
\newblock \showarticletitle{Automata for Reduction Properties Solving}.
\newblock \bibinfo{journal}{\emph{Journal of Symbolic Computation}}
  \bibinfo{volume}{20}, \bibinfo{number}{2} (\bibinfo{year}{1995}),
  \bibinfo{pages}{215--233}.
\newblock
\showISSN{0747-7171}
\urldef\tempurl%
\url{https://doi.org/10.1006/jsco.1995.1048}
\showDOI{\tempurl}


\bibitem[de~Moura and Bj{\o}rner(2008)]%
        {z3}
\bibfield{author}{\bibinfo{person}{Leonardo de Moura} {and}
  \bibinfo{person}{Nikolaj Bj{\o}rner}.} \bibinfo{year}{2008}\natexlab{}.
\newblock \showarticletitle{Z3: An Efficient SMT Solver}. In
  \bibinfo{booktitle}{\emph{Tools and Algorithms for the Construction and
  Analysis of Systems}}, \bibfield{editor}{\bibinfo{person}{C.~R. Ramakrishnan}
  {and} \bibinfo{person}{Jakob Rehof}} (Eds.). \bibinfo{publisher}{Springer
  Berlin Heidelberg}, \bibinfo{address}{Berlin, Heidelberg},
  \bibinfo{pages}{337--340}.
\newblock
\showISBNx{978-3-540-78800-3}


\bibitem[Feng et~al\mbox{.}(2018)]%
        {cdcl-synthesis}
\bibfield{author}{\bibinfo{person}{Yu Feng}, \bibinfo{person}{Ruben Martins},
  \bibinfo{person}{Osbert Bastani}, {and} \bibinfo{person}{Isil Dillig}.}
  \bibinfo{year}{2018}\natexlab{}.
\newblock \showarticletitle{Program Synthesis Using Conflict-Driven Learning}.
  In \bibinfo{booktitle}{\emph{Proceedings of the 39th ACM SIGPLAN Conference
  on Programming Language Design and Implementation}} (Philadelphia, PA, USA)
  \emph{(\bibinfo{series}{PLDI 2018})}. \bibinfo{publisher}{Association for
  Computing Machinery}, \bibinfo{address}{New York, NY, USA},
  \bibinfo{pages}{420–435}.
\newblock
\showISBNx{9781450356985}
\urldef\tempurl%
\url{https://doi.org/10.1145/3192366.3192382}
\showDOI{\tempurl}


\bibitem[Feng et~al\mbox{.}(2017a)]%
        {table-synthesis}
\bibfield{author}{\bibinfo{person}{Yu Feng}, \bibinfo{person}{Ruben Martins},
  \bibinfo{person}{Jacob Van~Geffen}, \bibinfo{person}{Isil Dillig}, {and}
  \bibinfo{person}{Swarat Chaudhuri}.} \bibinfo{year}{2017}\natexlab{a}.
\newblock \showarticletitle{Component-Based Synthesis of Table Consolidation
  and Transformation Tasks from Examples}. In
  \bibinfo{booktitle}{\emph{Proceedings of the 38th ACM SIGPLAN Conference on
  Programming Language Design and Implementation}} (Barcelona, Spain)
  \emph{(\bibinfo{series}{PLDI 2017})}. \bibinfo{publisher}{Association for
  Computing Machinery}, \bibinfo{address}{New York, NY, USA},
  \bibinfo{pages}{422–436}.
\newblock
\showISBNx{9781450349888}
\urldef\tempurl%
\url{https://doi.org/10.1145/3062341.3062351}
\showDOI{\tempurl}


\bibitem[Feng et~al\mbox{.}(2017b)]%
        {sypet}
\bibfield{author}{\bibinfo{person}{Yu Feng}, \bibinfo{person}{Ruben Martins},
  \bibinfo{person}{Yuepeng Wang}, \bibinfo{person}{Isil Dillig}, {and}
  \bibinfo{person}{Thomas~W. Reps}.} \bibinfo{year}{2017}\natexlab{b}.
\newblock \showarticletitle{Component-Based Synthesis for Complex APIs}. In
  \bibinfo{booktitle}{\emph{Proceedings of the 44th ACM SIGPLAN Symposium on
  Principles of Programming Languages}} (Paris, France)
  \emph{(\bibinfo{series}{POPL 2017})}. \bibinfo{publisher}{Association for
  Computing Machinery}, \bibinfo{address}{New York, NY, USA},
  \bibinfo{pages}{599–612}.
\newblock
\showISBNx{9781450346603}
\urldef\tempurl%
\url{https://doi.org/10.1145/3009837.3009851}
\showDOI{\tempurl}


\bibitem[Feser et~al\mbox{.}(2023)]%
        {Dillig23}
\bibfield{author}{\bibinfo{person}{Jack Feser}, \bibinfo{person}{I\c{s}\i{}l
  Dillig}, {and} \bibinfo{person}{Armando Solar-Lezama}.}
  \bibinfo{year}{2023}\natexlab{}.
\newblock \showarticletitle{Inductive Program Synthesis Guided by Observational
  Program Similarity}.
\newblock \bibinfo{journal}{\emph{Proc. ACM Program. Lang.}}
  \bibinfo{volume}{7}, \bibinfo{number}{OOPSLA2}, Article
  \bibinfo{articleno}{254} (\bibinfo{date}{oct} \bibinfo{year}{2023}),
  \bibinfo{numpages}{29}~pages.
\newblock
\urldef\tempurl%
\url{https://doi.org/10.1145/3622830}
\showDOI{\tempurl}


\bibitem[Finkbeiner et~al\mbox{.}(2019)]%
        {reactive}
\bibfield{author}{\bibinfo{person}{Bernd Finkbeiner}, \bibinfo{person}{Felix
  Klein}, \bibinfo{person}{Ruzica Piskac}, {and} \bibinfo{person}{Mark
  Santolucito}.} \bibinfo{year}{2019}\natexlab{}.
\newblock \showarticletitle{Synthesizing Functional Reactive Programs}. In
  \bibinfo{booktitle}{\emph{Proceedings of the 12th ACM SIGPLAN International
  Symposium on Haskell}} (Berlin, Germany) \emph{(\bibinfo{series}{Haskell
  2019})}. \bibinfo{publisher}{Association for Computing Machinery},
  \bibinfo{address}{New York, NY, USA}, \bibinfo{pages}{162–175}.
\newblock
\showISBNx{9781450368131}
\urldef\tempurl%
\url{https://doi.org/10.1145/3331545.3342601}
\showDOI{\tempurl}


\bibitem[Flanagan et~al\mbox{.}(1993)]%
        {flanagan}
\bibfield{author}{\bibinfo{person}{Cormac Flanagan}, \bibinfo{person}{Amr
  Sabry}, \bibinfo{person}{Bruce~F. Duba}, {and} \bibinfo{person}{Matthias
  Felleisen}.} \bibinfo{year}{1993}\natexlab{}.
\newblock \showarticletitle{The Essence of Compiling with Continuations}. In
  \bibinfo{booktitle}{\emph{Proceedings of the ACM SIGPLAN 1993 Conference on
  Programming Language Design and Implementation}} (Albuquerque, New Mexico,
  USA) \emph{(\bibinfo{series}{PLDI '93})}. \bibinfo{publisher}{Association for
  Computing Machinery}, \bibinfo{address}{New York, NY, USA},
  \bibinfo{pages}{237–247}.
\newblock
\showISBNx{0897915984}
\urldef\tempurl%
\url{https://doi.org/10.1145/155090.155113}
\showDOI{\tempurl}


\bibitem[Furcy and Koenig(2005)]%
        {10.5555/1642293.1642313}
\bibfield{author}{\bibinfo{person}{David Furcy} {and} \bibinfo{person}{Sven
  Koenig}.} \bibinfo{year}{2005}\natexlab{}.
\newblock \showarticletitle{Limited Discrepancy Beam Search}. In
  \bibinfo{booktitle}{\emph{Proceedings of the 19th International Joint
  Conference on Artificial Intelligence}} (Edinburgh, Scotland)
  \emph{(\bibinfo{series}{IJCAI'05})}. \bibinfo{publisher}{Morgan Kaufmann
  Publishers Inc.}, \bibinfo{address}{San Francisco, CA, USA},
  \bibinfo{pages}{125–131}.
\newblock


\bibitem[Guo et~al\mbox{.}(2019)]%
        {tygus}
\bibfield{author}{\bibinfo{person}{Zheng Guo}, \bibinfo{person}{Michael James},
  \bibinfo{person}{David Justo}, \bibinfo{person}{Jiaxiao Zhou},
  \bibinfo{person}{Ziteng Wang}, \bibinfo{person}{Ranjit Jhala}, {and}
  \bibinfo{person}{Nadia Polikarpova}.} \bibinfo{year}{2019}\natexlab{}.
\newblock \showarticletitle{Program Synthesis by Type-Guided Abstraction
  Refinement}.
\newblock \bibinfo{journal}{\emph{Proc. ACM Program. Lang.}}
  \bibinfo{volume}{4}, \bibinfo{number}{POPL}, Article \bibinfo{articleno}{12}
  (\bibinfo{date}{Dec.} \bibinfo{year}{2019}), \bibinfo{numpages}{28}~pages.
\newblock
\urldef\tempurl%
\url{https://doi.org/10.1145/3371080}
\showDOI{\tempurl}


\bibitem[Guria et~al\mbox{.}(2021)]%
        {rbsyn}
\bibfield{author}{\bibinfo{person}{Sankha~Narayan Guria},
  \bibinfo{person}{Jeffrey~S. Foster}, {and} \bibinfo{person}{David Van~Horn}.}
  \bibinfo{year}{2021}\natexlab{}.
\newblock \showarticletitle{RbSyn: Type- and Effect-Guided Program Synthesis}.
  In \bibinfo{booktitle}{\emph{Proceedings of the 42nd ACM SIGPLAN
  International Conference on Programming Language Design and Implementation}}
  (Virtual, Canada) \emph{(\bibinfo{series}{PLDI 2021})}.
  \bibinfo{publisher}{Association for Computing Machinery},
  \bibinfo{address}{New York, NY, USA}, \bibinfo{pages}{344–358}.
\newblock
\showISBNx{9781450383912}
\urldef\tempurl%
\url{https://doi.org/10.1145/3453483.3454048}
\showDOI{\tempurl}


\bibitem[Itzhaky et~al\mbox{.}(2017)]%
        {database-examples}
\bibfield{author}{\bibinfo{person}{Shachar Itzhaky}, \bibinfo{person}{Tomer
  Kotek}, \bibinfo{person}{Noam Rinetzky}, \bibinfo{person}{Mooly Sagiv},
  \bibinfo{person}{Orr Tamir}, \bibinfo{person}{Helmut Veith}, {and}
  \bibinfo{person}{Florian Zuleger}.} \bibinfo{year}{2017}\natexlab{}.
\newblock \showarticletitle{On the Automated Verification of Web Applications
  with Embedded {SQL}}. In \bibinfo{booktitle}{\emph{20th International
  Conference on Database Theory, {ICDT} 2017, March 21-24, 2017, Venice,
  Italy}} \emph{(\bibinfo{series}{LIPIcs}, Vol.~\bibinfo{volume}{68})},
  \bibfield{editor}{\bibinfo{person}{Michael Benedikt} {and}
  \bibinfo{person}{Giorgio Orsi}} (Eds.). \bibinfo{publisher}{Schloss Dagstuhl
  - Leibniz-Zentrum f{\"{u}}r Informatik}, \bibinfo{pages}{16:1--16:18}.
\newblock
\urldef\tempurl%
\url{https://doi.org/10.4230/LIPIcs.ICDT.2017.16}
\showDOI{\tempurl}


\bibitem[James et~al\mbox{.}(2020)]%
        {digging-fold}
\bibfield{author}{\bibinfo{person}{Michael~B. James}, \bibinfo{person}{Zheng
  Guo}, \bibinfo{person}{Ziteng Wang}, \bibinfo{person}{Shivani Doshi},
  \bibinfo{person}{Hila Peleg}, \bibinfo{person}{Ranjit Jhala}, {and}
  \bibinfo{person}{Nadia Polikarpova}.} \bibinfo{year}{2020}\natexlab{}.
\newblock \showarticletitle{Digging for Fold: Synthesis-Aided API Discovery for
  Haskell}.
\newblock \bibinfo{journal}{\emph{Proc. ACM Program. Lang.}}
  \bibinfo{volume}{4}, \bibinfo{number}{OOPSLA}, Article
  \bibinfo{articleno}{205} (\bibinfo{date}{nov} \bibinfo{year}{2020}),
  \bibinfo{numpages}{27}~pages.
\newblock
\urldef\tempurl%
\url{https://doi.org/10.1145/3428273}
\showDOI{\tempurl}

\bibitem[James et~al\mbox{.}(2020)]%
        {hoogleplus}
\bibfield{author}{\bibinfo{person}{Michael B. James}, \bibinfo{person}{Zheng
  Guo}, \bibinfo{person}{Ziteng Wang}, \bibinfo{person}{Shivani Doshi},
  \bibinfo{person}{Hila Peleg}, \bibinfo{person}{Ranjit Jhala}, {and}
  \bibinfo{person}{Nadia Polikarpova}.}
  \bibinfo{year}{2020}\natexlab{}.
\newblock \showarticletitle{Digging for Fold: Synthesis-Aided API Discovery for
  Haskell}.
\newblock \bibinfo{journal}{\emph{Proc. ACM Program. Lang.}} \bibinfo{volume}{4},
   \bibinfo{number}{OOPSLA} (\bibinfo{date}{Nov.} \bibinfo{year}{2020}),
  \bibinfo{articleno}{205}, \bibinfo{numpages}{27}~pages.
\newblock
\showISSN{2475-1421}
\urldef\tempurl%
\url{https://doi.org/10.1145/3428273}
\showDOI{\tempurl}

\bibitem[Jha et~al\mbox{.}(2010)]%
        {oracle-guided-synthesis}
\bibfield{author}{\bibinfo{person}{Susmit Jha}, \bibinfo{person}{Sumit
  Gulwani}, \bibinfo{person}{Sanjit~A. Seshia}, {and} \bibinfo{person}{Ashish
  Tiwari}.} \bibinfo{year}{2010}\natexlab{}.
\newblock \showarticletitle{Oracle-Guided Component-Based Program Synthesis}.
  In \bibinfo{booktitle}{\emph{Proceedings of the 32nd ACM/IEEE International
  Conference on Software Engineering - Volume 1}} (Cape Town, South Africa)
  \emph{(\bibinfo{series}{ICSE '10})}. \bibinfo{publisher}{Association for
  Computing Machinery}, \bibinfo{address}{New York, NY, USA},
  \bibinfo{pages}{215–224}.
\newblock
\showISBNx{9781605587196}
\urldef\tempurl%
\url{https://doi.org/10.1145/1806799.1806833}
\showDOI{\tempurl}


\bibitem[Jhala and Vazou(2021)]%
        {JV21}
\bibfield{author}{\bibinfo{person}{Ranjit Jhala} {and} \bibinfo{person}{Niki
  Vazou}.} \bibinfo{year}{2021}\natexlab{}.
\newblock \showarticletitle{{Refinement Types: {A} Tutorial}}.
\newblock \bibinfo{journal}{\emph{Found. Trends Program. Lang.}}
  \bibinfo{volume}{6}, \bibinfo{number}{3-4} (\bibinfo{year}{2021}),
  \bibinfo{pages}{159--317}.
\newblock
\urldef\tempurl%
\url{https://doi.org/10.1561/2500000032}
\showDOI{\tempurl}


\bibitem[Koppel et~al\mbox{.}(2022)]%
        {ecta}
\bibfield{author}{\bibinfo{person}{James Koppel}, \bibinfo{person}{Zheng Guo},
  \bibinfo{person}{Edsko de Vries}, \bibinfo{person}{Armando Solar-Lezama},
  {and} \bibinfo{person}{Nadia Polikarpova}.} \bibinfo{year}{2022}\natexlab{}.
\newblock \showarticletitle{Searching Entangled Program Spaces}.
\newblock \bibinfo{journal}{\emph{Proc. ACM Program. Lang.}}
  \bibinfo{volume}{6}, \bibinfo{number}{ICFP}, Article \bibinfo{articleno}{91}
  (\bibinfo{date}{aug} \bibinfo{year}{2022}), \bibinfo{numpages}{29}~pages.
\newblock
\urldef\tempurl%
\url{https://doi.org/10.1145/3547622}
\showDOI{\tempurl}


\bibitem[Leroy et~al\mbox{.}(2022)]%
        {ocaml}
\bibfield{author}{\bibinfo{person}{Xavier Leroy},
  \bibinfo{person}{Didier~R\'emy Alain~Frisch, Jacques~Garrigue}, {and}
  \bibinfo{person}{J\'er{\^o}me Vouillon}.} \bibinfo{year}{2022}\natexlab{}.
\newblock \bibinfo{title}{Parsing with Ocamllex}.
\newblock
\newblock
\urldef\tempurl%
\url{https://ocaml.org/manual/lexyacc.html}
\showURL{%
\tempurl}


\bibitem[Miltner et~al\mbox{.}(2022)]%
        {Miltner2022}
\bibfield{author}{\bibinfo{person}{Anders Miltner},
  \bibinfo{person}{Adrian~Trejo Nu\~{n}ez}, \bibinfo{person}{Ana Brendel},
  \bibinfo{person}{Swarat Chaudhuri}, {and} \bibinfo{person}{Isil Dillig}.}
  \bibinfo{year}{2022}\natexlab{}.
\newblock \showarticletitle{Bottom-up Synthesis of Recursive Functional
  Programs Using Angelic Execution}.
\newblock \bibinfo{journal}{\emph{Proc. ACM Program. Lang.}}
  \bibinfo{volume}{6}, \bibinfo{number}{POPL}, Article \bibinfo{articleno}{21}
  (\bibinfo{date}{jan} \bibinfo{year}{2022}), \bibinfo{numpages}{29}~pages.
\newblock
\urldef\tempurl%
\url{https://doi.org/10.1145/3498682}
\showDOI{\tempurl}


\bibitem[Mishra and Jagannathan(2022)]%
        {cobalt-tech}
\bibfield{author}{\bibinfo{person}{Ashish Mishra} {and} \bibinfo{person}{Suresh
  Jagannathan}.} \bibinfo{year}{2022}\natexlab{}.
\newblock \showarticletitle{Specification-Guided Component-Based Synthesis from
  Effectful Libraries}.
\newblock \bibinfo{journal}{\emph{Proc. ACM Program. Lang.}}
  \bibinfo{volume}{6}, \bibinfo{number}{OOPSLA2}, Article
  \bibinfo{articleno}{147} (\bibinfo{date}{oct} \bibinfo{year}{2022}),
  \bibinfo{numpages}{30}~pages.
\newblock
\urldef\tempurl%
\url{https://doi.org/10.1145/3563310}
\showDOI{\tempurl}


\bibitem[Polikarpova et~al\mbox{.}(2016)]%
        {synquid}
\bibfield{author}{\bibinfo{person}{Nadia Polikarpova}, \bibinfo{person}{Ivan
  Kuraj}, {and} \bibinfo{person}{Armando Solar-Lezama}.}
  \bibinfo{year}{2016}\natexlab{}.
\newblock \showarticletitle{Program Synthesis from Polymorphic Refinement
  Types}. In \bibinfo{booktitle}{\emph{Proceedings of the 37th ACM SIGPLAN
  Conference on Programming Language Design and Implementation}} (Santa
  Barbara, CA, USA) \emph{(\bibinfo{series}{PLDI '16})}.
  \bibinfo{publisher}{Association for Computing Machinery},
  \bibinfo{address}{New York, NY, USA}, \bibinfo{pages}{522–538}.
\newblock
\showISBNx{9781450342612}
\urldef\tempurl%
\url{https://doi.org/10.1145/2908080.2908093}
\showDOI{\tempurl}


\bibitem[Polozov and Gulwani(2015)]%
        {flashfill15}
\bibfield{author}{\bibinfo{person}{Oleksandr Polozov} {and}
  \bibinfo{person}{Sumit Gulwani}.} \bibinfo{year}{2015}\natexlab{}.
\newblock \showarticletitle{FlashMeta: A Framework for Inductive Program
  Synthesis}. In \bibinfo{booktitle}{\emph{Proceedings of the 2015 ACM SIGPLAN
  International Conference on Object-Oriented Programming, Systems, Languages,
  and Applications}} (Pittsburgh, PA, USA) \emph{(\bibinfo{series}{OOPSLA
  2015})}. \bibinfo{publisher}{Association for Computing Machinery},
  \bibinfo{address}{New York, NY, USA}, \bibinfo{pages}{107–126}.
\newblock
\showISBNx{9781450336895}
\urldef\tempurl%
\url{https://doi.org/10.1145/2814270.2814310}
\showDOI{\tempurl}


\bibitem[Shi et~al\mbox{.}(2019)]%
        {frangel}
\bibfield{author}{\bibinfo{person}{Kensen Shi}, \bibinfo{person}{Jacob
  Steinhardt}, {and} \bibinfo{person}{Percy Liang}.}
  \bibinfo{year}{2019}\natexlab{}.
\newblock \showarticletitle{FrAngel: Component-Based Synthesis with Control
  Structures}.
\newblock \bibinfo{journal}{\emph{Proc. ACM Program. Lang.}}
  \bibinfo{volume}{3}, \bibinfo{number}{POPL}, Article \bibinfo{articleno}{73}
  (\bibinfo{date}{jan} \bibinfo{year}{2019}), \bibinfo{numpages}{29}~pages.
\newblock
\urldef\tempurl%
\url{https://doi.org/10.1145/3290386}
\showDOI{\tempurl}


\bibitem[Swamy et~al\mbox{.}(2013)]%
        {fstar}
\bibfield{author}{\bibinfo{person}{Nikhil Swamy}, \bibinfo{person}{Joel
  Weinberger}, \bibinfo{person}{Cole Schlesinger}, \bibinfo{person}{Juan Chen},
  {and} \bibinfo{person}{Benjamin Livshits}.} \bibinfo{year}{2013}\natexlab{}.
\newblock \showarticletitle{{Verifying Higher-Order Programs with the Dijkstra
  Monad}}. In \bibinfo{booktitle}{\emph{Proceedings of the 34th ACM SIGPLAN
  Conference on Programming Language Design and Implementation}} (Seattle,
  Washington, USA) \emph{(\bibinfo{series}{PLDI '13})}.
  \bibinfo{publisher}{Association for Computing Machinery},
  \bibinfo{address}{New York, NY, USA}, \bibinfo{pages}{387–398}.
\newblock
\showISBNx{9781450320146}
\urldef\tempurl%
\url{https://doi.org/10.1145/2491956.2491978}
\showDOI{\tempurl}


\bibitem[Vazou et~al\mbox{.}(2015)]%
        {liquidextended}
\bibfield{author}{\bibinfo{person}{Niki Vazou}, \bibinfo{person}{Alexander
  Bakst}, {and} \bibinfo{person}{Ranjit Jhala}.}
  \bibinfo{year}{2015}\natexlab{}.
\newblock \showarticletitle{Bounded Refinement Types}. In
  \bibinfo{booktitle}{\emph{Proceedings of the 20th ACM SIGPLAN International
  Conference on Functional Programming}} (Vancouver, BC, Canada)
  \emph{(\bibinfo{series}{ICFP 2015})}. \bibinfo{publisher}{Association for
  Computing Machinery}, \bibinfo{address}{New York, NY, USA},
  \bibinfo{pages}{48--61}.
\newblock
\showISBNx{9781450336697}
\urldef\tempurl%
\url{https://doi.org/10.1145/2784731.2784745}
\showDOI{\tempurl}

\bibitem[Rondon et~al\mbox{.}(2008)]%
        {liquidoriginal}
\bibfield{author}{\bibinfo{person}{Patrick M. Rondon}, \bibinfo{person}{Ming
  Kawaguci}, {and} \bibinfo{person}{Ranjit Jhala}.}
  \bibinfo{year}{2008}\natexlab{}.
\newblock \showarticletitle{Liquid Types}. In
  \bibinfo{booktitle}{\emph{Proceedings of the 29th ACM SIGPLAN Conference on
  Programming Language Design and Implementation}} (Tucson, AZ, USA)
  \emph{(\bibinfo{series}{PLDI '08})}. \bibinfo{publisher}{Association for
  Computing Machinery}, \bibinfo{address}{New York, NY, USA},
  \bibinfo{pages}{159--169}.
\newblock
\showISBNx{9781595938602}
\urldef\tempurl%
\url{https://doi.org/10.1145/1375581.1375602}
\showDOI{\tempurl}


\bibitem[Lau et~al\mbox{.}(2000)]%
        {vsa}
\bibfield{author}{\bibinfo{person}{Tessa A. Lau}, \bibinfo{person}{Pedro
  Domingos}, {and} \bibinfo{person}{Daniel S. Weld}.}
  \bibinfo{year}{2000}\natexlab{}.
\newblock \showarticletitle{Version Space Algebra and Its Application to
  Programming by Demonstration}. In \bibinfo{booktitle}{\emph{Proceedings of the
  Seventeenth International Conference on Machine Learning}}
  \emph{(\bibinfo{series}{ICML '00'})}. \bibinfo{publisher}{Morgan Kaufmann
  Publishers Inc.}, \bibinfo{address}{San Francisco, CA, USA},
  \bibinfo{pages}{527--534}.
\newblock
\showISBNx{1558607072}


\bibitem[Wang et~al\mbox{.}(2019)]%
        {viser}
\bibfield{author}{\bibinfo{person}{Chenglong Wang}, \bibinfo{person}{Yu Feng},
  \bibinfo{person}{Rastislav Bodik}, \bibinfo{person}{Alvin Cheung}, {and}
  \bibinfo{person}{Isil Dillig}.} \bibinfo{year}{2019}\natexlab{}.
\newblock \showarticletitle{Visualization by Example}.
\newblock \bibinfo{journal}{\emph{Proc. ACM Program. Lang.}}
  \bibinfo{volume}{4}, \bibinfo{number}{POPL}, Article \bibinfo{articleno}{49}
  (\bibinfo{date}{dec} \bibinfo{year}{2019}), \bibinfo{numpages}{28}~pages.
\newblock
\urldef\tempurl%
\url{https://doi.org/10.1145/3371117}
\showDOI{\tempurl}


\bibitem[Wang et~al\mbox{.}(2017)]%
        {dace}
\bibfield{author}{\bibinfo{person}{Xinyu Wang}, \bibinfo{person}{Isil Dillig},
  {and} \bibinfo{person}{Rishabh Singh}.} \bibinfo{year}{2017}\natexlab{}.
\newblock \showarticletitle{Synthesis of Data Completion Scripts Using Finite
  Tree Automata}.
\newblock \bibinfo{journal}{\emph{Proc. ACM Program. Lang.}}
  \bibinfo{volume}{1}, \bibinfo{number}{OOPSLA}, Article
  \bibinfo{articleno}{62} (\bibinfo{date}{oct} \bibinfo{year}{2017}),
  \bibinfo{numpages}{26}~pages.
\newblock
\urldef\tempurl%
\url{https://doi.org/10.1145/3133886}
\showDOI{\tempurl}


\bibitem[Willsey et~al\mbox{.}(2021)]%
        {Egg}
\bibfield{author}{\bibinfo{person}{Max Willsey}, \bibinfo{person}{Chandrakana
  Nandi}, \bibinfo{person}{Yisu~Remy Wang}, \bibinfo{person}{Oliver Flatt},
  \bibinfo{person}{Zachary Tatlock}, {and} \bibinfo{person}{Pavel Panchekha}.}
  \bibinfo{year}{2021}\natexlab{}.
\newblock \showarticletitle{Egg: Fast and Extensible Equality Saturation}.
\newblock \bibinfo{journal}{\emph{Proc. ACM Program. Lang.}}
  \bibinfo{volume}{5}, \bibinfo{number}{POPL}, Article \bibinfo{articleno}{23}
  (\bibinfo{date}{jan} \bibinfo{year}{2021}), \bibinfo{numpages}{29}~pages.
\newblock
\urldef\tempurl%
\url{https://doi.org/10.1145/3434304}
\showDOI{\tempurl}


\bibitem[Wolfman et~al\mbox{.}(2001)]%
        {VSA03}
\bibfield{author}{\bibinfo{person}{Steven Wolfman}, \bibinfo{person}{Pedro
  Domingos}, {and} \bibinfo{person}{Daniel Weld}.}
  \bibinfo{year}{2001}\natexlab{}.
\newblock \showarticletitle{Programming By Demonstration Using Version Space
  Algebra}.
\newblock \bibinfo{journal}{\emph{Machine Learning}}  \bibinfo{volume}{53}
  (\bibinfo{date}{12} \bibinfo{year}{2001}).
\newblock
\urldef\tempurl%
\url{https://doi.org/10.1023/A:1025671410623}
\showDOI{\tempurl}


\bibitem[Yuan et~al\mbox{.}(2023)]%
        {YRS23}
\bibfield{author}{\bibinfo{person}{Yongwei Yuan}, \bibinfo{person}{Arjun
  Radhakrishna}, {and} \bibinfo{person}{Roopsha Samanta}.}
  \bibinfo{year}{2023}\natexlab{}.
\newblock \showarticletitle{{Trace-Guided Inductive Synthesis of Recursive
  Functional Programs}}.
\newblock  \bibinfo{volume}{7}, \bibinfo{number}{PLDI} (\bibinfo{year}{2023}).
\newblock
\urldef\tempurl%
\url{https://doi.org/10.1145/3591255}
\showDOI{\tempurl}


\bibitem[Zhang et~al\mbox{.}(2023)]%
        {egg-analysis}
\bibfield{author}{\bibinfo{person}{Yihong Zhang}, \bibinfo{person}{Yisu~Remy
  Wang}, \bibinfo{person}{Oliver Flatt}, \bibinfo{person}{David Cao},
  \bibinfo{person}{Philip Zucker}, \bibinfo{person}{Eli Rosenthal},
  \bibinfo{person}{Zachary Tatlock}, {and} \bibinfo{person}{Max Willsey}.}
  \bibinfo{year}{2023}\natexlab{}.
\newblock \showarticletitle{Better Together: Unifying Datalog and Equality
  Saturation}.
\newblock \bibinfo{journal}{\emph{Proc. ACM Program. Lang.}}
  \bibinfo{volume}{7}, \bibinfo{number}{PLDI}, Article \bibinfo{articleno}{125}
  (\bibinfo{date}{jun} \bibinfo{year}{2023}), \bibinfo{numpages}{25}~pages.
\newblock
\urldef\tempurl%
\url{https://doi.org/10.1145/3591239}
\showDOI{\tempurl}

\bibitem[Miltner et~al\mbox{.}(2024)]%
        {cata}
\bibfield{author}{\bibinfo{person}{Anders Miltner}, \bibinfo{person}{Ziteng
  Wang}, \bibinfo{person}{Swarat Chaudhuri}, {and} \bibinfo{person}{Isil
  Dillig}.}
  \bibinfo{year}{2024}\natexlab{}.
\newblock \showarticletitle{Relational Synthesis of Recursive Programs via
  Constraint Annotated Tree Automata}. In \bibinfo{booktitle}{\emph{Computer
  Aided Verification: 36th International Conference, CAV 2024, Montreal, QC,
  Canada, July 24--27, 2024, Proceedings, Part III}} (Montreal, QC, Canada).
  \bibinfo{publisher}{Springer-Verlag}, \bibinfo{address}{Berlin, Heidelberg},
  \bibinfo{pages}{41--63}.
\newblock
\showISBNx{978-3-031-65632-3}
\urldef\tempurl%
\url{https://doi.org/10.1007/978-3-031-65633-0_3}
\showDOI{\tempurl}

\bibitem[Kaki and Jagannathan(2014)]%
        {relref}
\bibfield{author}{\bibinfo{person}{Gowtham Kaki} {and} \bibinfo{person}{Suresh
  Jagannathan}.}
  \bibinfo{year}{2014}\natexlab{}.
\newblock \showarticletitle{A Relational Framework for Higher-Order Shape
  Analysis}. In \bibinfo{booktitle}{\emph{Proceedings of the 19th ACM SIGPLAN
  International Conference on Functional Programming}} (Gothenburg, Sweden)
  \emph{(\bibinfo{series}{ICFP '14})}. \bibinfo{publisher}{Association for
  Computing Machinery}, \bibinfo{address}{New York, NY, USA},
  \bibinfo{pages}{311--324}.
\newblock
\showISBNx{9781450328739}
\urldef\tempurl%
\url{https://doi.org/10.1145/2628136.2628159}
\showDOI{\tempurl}

\bibitem[Smith and Albarghouthi(2019)]%
        {manual}
\bibfield{author}{\bibinfo{person}{Calvin Smith} {and} \bibinfo{person}{Aws
  Albarghouthi}.} \bibinfo{year}{2019}\natexlab{}.
\newblock \showarticletitle{Program Synthesis with Equivalence Reduction}.
\newblock In \bibinfo{booktitle}{\emph{Verification, Model Checking, and
  Abstract Interpretation}}, \bibfield{editor}{\bibinfo{person}{Constantin
  Enea} {and} \bibinfo{person}{Ruzica Piskac}} (Eds.). \bibinfo{publisher}{Springer International Publishing},
  \bibinfo{address}{Cham}, \bibinfo{pages}{24--47}.
\newblock
\showISBNx{978-3-030-11245-5}
\urldef\tempurl%
\url{https://doi.org/10.1007/978-3-030-11245-5_2}
\showDOI{\tempurl}

\bibitem[Wang et~al\mbox{.}(2017)]%
        {blaze}
\bibfield{author}{\bibinfo{person}{Xinyu Wang}, \bibinfo{person}{Isil Dillig},
  {and} \bibinfo{person}{Rishabh Singh}.} \bibinfo{year}{2017}\natexlab{}.
\newblock \showarticletitle{Program Synthesis using Abstraction Refinement}.
\newblock \bibinfo{journal}{\emph{Proc. ACM Program. Lang.}} \bibinfo{volume}{2},
  \bibinfo{number}{POPL} (\bibinfo{date}{Dec.} \bibinfo{year}{2017}),
  \bibinfo{articleno}{63}, \bibinfo{numpages}{30}~pages.
\newblock
\showISSN{2475-1421}
\urldef\tempurl%
\url{https://doi.org/10.1145/3158151}
\showDOI{\tempurl}

\bibitem[Nielson et~al\mbox{.}(1999)]%
        {neilson}
\bibfield{author}{\bibinfo{person}{Flemming Nielson},
  \bibinfo{person}{Hanne Riis Nielson}, {and} \bibinfo{person}{Chris Hankin}.}
  \bibinfo{year}{1999}\natexlab{}.
\newblock \showarticletitle{Abstract Interpretation}.
\newblock In \bibinfo{booktitle}{\emph{Principles of Program Analysis}}.
  \bibinfo{publisher}{Springer Berlin Heidelberg}, \bibinfo{address}{Berlin,
  Heidelberg}, \bibinfo{pages}{211--282}.
\newblock
\urldef\tempurl%
\url{https://doi.org/10.1007/978-3-662-03811-6_4}
\showDOI{\tempurl}
\showISBNx{978-3-662-03811-6}

\end{thebibliography}

%% \begin{acks}
%% To Robert, for the bagels and explaining CMYK and color spaces.
%% \end{acks}

%% \bibliographystyle{ACM-Reference-Format}
%% \bibliography{sample-base}

%\input{appendix}

\end{document}